\documentclass[aps, prd, twocolumn, amsmath, floats, floatfix, superscriptaddress, nofootinbib]{revtex4-2}

\usepackage{graphicx}
\usepackage{epsfig}
\usepackage{amsmath,amsfonts,amssymb}
\usepackage{caption}
\usepackage{subcaption}
\usepackage{wrapfig}
\usepackage{verbatim}
\usepackage{float}
\usepackage{morefloats}
\usepackage[dvipsnames]{xcolor}
\usepackage{booktabs}
\usepackage[normalem]{ulem}
\usepackage{multirow}
\usepackage{makecell}
\usepackage{tabulary}
\usepackage{cancel}
\usepackage{xcolor}
\usepackage{blindtext}      
\usepackage{hyperref}       
\hypersetup{
	colorlinks   = true, 
	urlcolor     = blue, 
	linkcolor    = blue, 
	citecolor    = red,  
}
\usepackage{url}

\usepackage{booktabs}       
\usepackage{multirow}       
\usepackage{adjustbox}      
\usepackage{cellspace}
\usepackage{tablefootnote}  
\usepackage{physics}                           
\usepackage{enumitem}
\setlist[itemize]{noitemsep, topsep=0cm, leftmargin = 1.2cm}        
\setlist[enumerate]{noitemsep, topsep=0cm, leftmargin = 1.2cm}      

\usepackage{ragged2e} 
\usepackage{subcaption} 
\DeclareCaptionJustification{justified}{\justifying} 

\begin{document}

\title{Descalarization bursts from Galactic supramassive neutron stars: a kHz continuous-wave target}
\author{Jos\'e Carlos Olvera M.}
\email{jose.olvera-meneses@uni-tuebingen.de}

\affiliation{Theoretical Astrophysics, IAAT, Eberhard Karls University of Tübingen, 72076 Tübingen, Germany}
\affiliation{Departamento de  Astronom\'{\i}a y Astrof\'{\i}sica, Universitat de Val\`encia,  Avinguda Vicent Andrés Estellés 19, 46100, Burjassot (Val\`encia), Spain}

\author{Daniela D. Doneva}
\affiliation{Theoretical Astrophysics, IAAT, Eberhard Karls University of Tübingen, 72076 Tübingen, Germany}
\affiliation{Departamento de  Astronom\'{\i}a y Astrof\'{\i}sica, Universitat de Val\`encia,  Avinguda Vicent Andrés Estellés 19, 46100, Burjassot (Val\`encia), Spain}

\author{Pablo Cerdá-Durán}
\affiliation{Departamento de  Astronom\'{\i}a y Astrof\'{\i}sica, Universitat de Val\`encia,  Avinguda Vicent Andrés Estellés 19, 46100, Burjassot (Val\`encia), Spain}
\affiliation{Observatorio Astron\'omico, Universitat de València, C/ Catedrático Jos\'e Beltr\'an 2, 46980, Paterna (València), Spain}

\author{José A. Font}
\affiliation{Departamento de  Astronom\'{\i}a y Astrof\'{\i}sica, Universitat de Val\`encia,  Avinguda Vicent Andrés Estellés 19, 46100, Burjassot (Val\`encia), Spain}
\affiliation{Observatorio Astron\'omico, Universitat de València, C/ Catedrático Jos\'e Beltr\'an 2, 46980, Paterna (València), Spain}

\author{Stoytcho S. Yazadjiev}
\affiliation{Department of Theoretical Physics, Faculty of Physics, Sofia University, Sofia 1164, Bulgaria}
\affiliation{Institute of Mathematics and Informatics, 	Bulgarian Academy of Sciences, 	Acad. G. Bonchev St. 8, Sofia 1113, Bulgaria}

            \begin{abstract}

                In massive scalar-tensor gravity, a scalarized neutron star must shed its scalar hair as it collapses into a black hole. Using 3D simulations of a realistic rotating progenitor we show that dispersion stretches the emitted burst into a quasi-monochromatic signal in the kHz frequency range lasting $>10^4$ yr and amenable to continuous-wave methods. We find that detecting these signals in directed searches of supernova remnants without a central pulsar requires sensitivity depths of 233--762 $\rm{Hz^{-1/2}}$ with the A+-ET-CE detector network. We estimate up to $\sim3$ of these signals in transit at any observational epoch.  
            \end{abstract}

\maketitle
\textbf{Introduction. } The most recent gravitational wave (GW) catalogs of the fourth observing run (O4) from the LIGO-Virgo-KAGRA (LVK) collaboration show no evidence for deviations from general relativity (GR)~\cite{TGR1,TGR2,TGR3,LIGOScientific:2026oim}.  Some extensions of GR with additional degrees of freedom, though, present only short-range deviations from GR, hence invisible to current tests. An example is  scalar-tensor theories (STT) \cite{Jordan1959,Brans1961}, which extend GR through a non-minimally coupled real scalar field \cite{Heisenberg2019}. In particular, massive STT (MSTT) endow the scalar field with a mass $m_\Phi$ restricting the scalar interaction to a scale $\lambda_\Phi=\hbar/m_\Phi c$. For high enough $m_\Phi$, deviations from GR become effectively invisible in Solar System ($\lambda_\Phi\lesssim 10^8$ km) \cite{Bertotti:2003rm,Will:2014kxa}  and binary pulsar ($\lambda_\Phi\lesssim 10^6$ km) \cite{Freire:2024adf, Zhao:2022vig} tests of gravity. Binary neutron star (NS) mergers reach separations of $\lesssim 30$ km, being the only system sensitive to short ranges and restrict $\lambda_\Phi\lesssim60$ km \cite{Kuan:2023hrh}. This short-range field still allows for significant effects in the strong-field regime of compact objects such as NS \cite{Ramazanolu2016,Yazadjiev:2016pcb}. 

STT allow for ``spontaneous scalarization'' \cite{Damour1993}, a phase-transition-like  phenomenon  \cite{Unluturk:2025zie} which endows a compact object with a non-trivial scalar field. For a NS, this effect depends on the equation of state (EoS) and the theory parameters \cite{Damour1993, Doneva2024}. STT with a real field respect a no hair theorem \cite{Yazadjiev:2025ezx}. This implies that the only stationary black hole (BH) solution in vacuum is the Kerr spacetime, inherently satisfying all BH tests of gravity. As a result, when a scalarized NS collapses, the scalar field is inevitably radiated away \cite{Novak1998,Kuan:2021lol,Lam2024,OlveraM:2026whf}.

MSTT can provide an additional scalar polarization mode with two components (breathing and longitudinal) that can be disentangled from the tensorial GW polarizations \cite{Nishizawa2009}. For the scalar emission, the scalar field mass induces a dispersive group velocity $v_g=\left[1-(f_\Phi/f)^2\right]^{1/2}$, stretching the scalar burst over time. 

The frequency band for this emission is set by $m_\Phi$. Modes of frequency $f$ are evanescent for $f\lesssim f_\Phi=m_\Phi c^2/h$. For a detector at a distance $D$ from the source, frequencies $f>f_\Phi$ arrive at $t(f)=D/v_g$, producing a long-lived, quasi-monochromatic signal. This was first studied by \cite{Sperhake:2017itk} in core-collapse simulations of non-rotating progenitors and  subsequently for light enough scalar masses to set the emission within the LVK band \cite{Rosca-Mead2020,Huang2021,Asakawa:2023obq,Rosca-Mead:2019seq,Cheong2019, Geng:2020slq}. Such masses are now excluded by GW170817 \cite{Kuan:2023hrh}. If both components of GW170817 were strongly scalarized, compatibility with observations requires $f_\Phi\gtrsim 2.4$ kHz. If only the primary component is, the bound relaxes to $f_\Phi\gtrsim800$ Hz \cite{Kuan:2023hrh}, both cases excluding emission in the most sensitive part of the LVK band. 

Post-merger searches in GW170817 and GW190425 have placed only upper limits in the 1--4 kHz band \cite{LIGOScientific:2017fdd}. Developments in detection capabilities will improved sensitivity and frequency band coverage \cite{LISA:2022kgy,ET_D_sensitivity,ET:2025xjr,CosmicExplorer2022}. Particularly, the post-merger tuned configurations of Cosmic Explorer (CE) will increase sensitivity in the 2-4 kHz band \cite{CosmicExplorer2022}.  Transient systems that radiate in this high-frequency band include post-merger remnants of NS mergers and core-collapse supernovae (CCSNe) \cite{Aggarwal:2025}. To our knowledge, descalarization bursts constitute the only long-lived kHz sources whose detectability improves by increasing observation time and for which the phase evolution is determined from the theory parameters and the remnant properties. Since the emission is a persistent signal, it is a target for continuous-wave (CW) methods \cite{Wette:2023dom,Riles:2022wwz}, often characterized by the sensitivity depth $\mathcal{D}$ achieved by a search \cite{Dreissigacker:2018afk,Behnke2015}, which measures the ratio of the detector network noise to the smallest detectable strain. Typical all-sky searches reach $\mathcal{D}\sim50\, \rm{Hz}^{-1/2}$ \cite{Steltner:2023cfk}  and cover frequencies below $\sim2$ kHz \cite{LVK2022,LIGOScientific:2026plm},  while targeted searches can reach up to $\mathcal{D}\sim 1000\,\rm{Hz^{-1/2}}$ \cite{Riles:2022wwz}. 

 Here, using 3D numerical-relativity simulations and scalar-field masses consistent with current bounds, we study the descalarization of a supramassive NS, born rapidly rotating, at the moment it is driven past the instability threshold as it spins down due to magnetic braking.  The MSTT parameters and EoS are fixed to admit stable configurations  at least as massive ($2.35\pm0.11M_\odot$) and rapidly rotating (707 Hz) as PSR~J0952$-$0607 \cite{Romani2022,Romani2025}. The progenitor we evolve lies on the same angular momentum sequence as PSR~J0952$-$0607, but on the unstable branch.  Including rotation in our models is important since it supports higher masses in STT and has been shown to enhance scalarization \cite{Doneva2013,OlveraM:2026whf}.  We find that the mass-induced dispersion of the emission yields a long-lived ($>10^4\rm{yr}$), quasi-monochromatic signal in the kHz range, whose frequency drifts at $|{\dot{f}}|< 10^{-8} \rm{\,Hz\, s^{-1}}$ and can be probed by CW searches.  We provide an upper-limit estimate of the number of signals present in the Galaxy at any given time and quantify for the first time the equivalent $\mathcal{D}$ needed to detect a descalarization burst. We argue that detection searches should be directed to Galactic SN remnants without a central pulsar (consistent with BH formation) and for which a depth within reach of existing directed searches \cite{Ming:2025ehy,Abac2026,Abbott2021_ContinuousGW_SNR} is achieved for a given set of theory parameters. 
 

\textbf{Massive scalar-tensor theories.} In MSTT, the action for a real scalar field non-minimally coupled to the geometric sector in the Jordan frame is \cite{Brans1961,Jordan1959}
\begin{equation}
\begin{aligned}
S_{\text{J}} =  \frac{1}{16\pi G} & \int d^4x \sqrt{-g}
\Big[  \Phi R    - \frac{\omega(\Phi)}{\Phi}
g^{\mu\nu} \nabla_\mu\Phi\nabla_\nu\Phi \\
 &- U(\Phi) \Big]  +\, S_{\text{matter}}(\Psi_{\text{matter}}, g_{\mu\nu})\,,
\end{aligned}
\label{eq:JFAction}
\end{equation}
where $R=g^{\mu\nu}R_{\mu\nu}$ is the curvature scalar for the Jordan frame metric $g_{\mu\nu}$, $G$ is bare gravitational constant and $\Phi$ is the scalar field. The term $U(\Phi)$ represents the potential of the scalar field, which includes its mass term, while $\omega(\Phi)$ is the coupling function that governs the kinetic term. The matter action $S_{\text{matter}}$ depends on the matter fields $\Psi_{\text{matter}}$, modeled here as a perfect fluid, and minimally coupled to the metric $g_{\mu\nu}$. 

The Jordan frame metric $g_{\mu\nu}$ is conformally related to the Einstein frame metric $g^*_{\mu\nu}$ through  $g^*_{\mu\nu} = A(\phi)^{-2}g_{\mu\nu}$, with $A(\phi)=\Phi^{-1/2}$, where $A(\phi)$ is the conformal factor and $\phi$ is the new field in the Einstein frame. 
We employ the Einstein frame coupling function proposed in \cite{Damour1993,Damour1992}, and potential $V(\phi)$ \begin{equation}\label{eq:coupling}
    A(\phi)= e^{\alpha_0\phi + \frac{1}{2}\beta_0\phi^2}, \quad V(\phi)= \frac{1}{2}m^2_{\Phi}\phi^2\,,
\end{equation}
where $\alpha_0$ and $\beta_0$ are the linear and quadratic coupling constants. GR is recovered in the limit $\alpha_0,\beta_0\rightarrow0$ while spontaneous scalarization is allowed for $\alpha_0=0$ and $\beta_0\neq0$ \cite{Doneva2024}. For scalar radiation to couple to the detector and produce additional polarizations  $\alpha_0\neq0$ is required \cite{Novak1998}. 

Our model is thus characterized by the parameters $(\alpha_0,\beta_0,m_{\Phi})$. Currently, GW170817 provides one of the most stringent constraints on $m_{\Phi}$ ~\cite{Kuan:2023hrh} within the parameter space allowed by Solar System experiments and  binary pulsar tests  \cite{Freire:2024adf,Batrakov2024,Bertotti:2003rm,Damour1996}. If the component stars were scalarized, the interaction between the scalar clouds, which sets in once their separation falls within a few Compton wavelengths, would drain orbital energy and produce a measurable dephasing during the inspiral phase \cite{Kuan:2023hrh}. Masses $m_\Phi\gtrsim 10^{-11} \rm{eV/c^2}$ suppress this effect, since the interaction only appears at the late inspiral, where it becomes degenerate with tidal effects. This is a conservative bound since it assumes that both stars are scalarized and it does not consider the degeneracies that may arise between the scalar charge and the binary parameters (chirp mass, mass ratio, spin) \cite{Sanger2026, Niu:2021nic}; no study has mapped such degeneracies yet.

If only one component of GW170817 were strongly scalarized, the asymmetry of scalar charge would source dipole radiation. Since dipole radiation is emitted only above orbital frequencies $f_{\rm orb}>f_{\Phi}$ (where $f_{\rm GW}= 2f_{\rm orb}$) \cite{Alsing:2011er}, lower scalar-field masses can remain compatible with the observational bound.  For the particular case of $m_\Phi=3.3\times10^{-12}\rm eV/c^2$, $f_\Phi\sim0.8$ kHz, dipole radiation is abscent below  $f_{\rm GW}=2f_{\Phi}\sim1.6\, \rm{kHz}$ and only appears above the frequency range where GW170817 accumulated most of its signal-to-noise ratio (SNR) \cite{LIGO2019_GW170817}.

We explore two scalar field masses representative for each case. The first one has a mass of $3.34\times10^{-12}$ eV$/c^2$, such that for suitable combination of $\beta_0$ and EoS, only stars with $M>1.36M_\odot$ are scalarized. In this case, the burst is above the cutoff frequency $f_\Phi \simeq 0.8$~kHz, within the sensitivity band of current and planned ground-based interferometers. For the second case,  $m_\Phi=10^{-11}$ eV$/c^2$, lies at the edge of the scalar-tidal degeneracy limit discussed in \cite{Kuan:2023hrh}. This mass allows only for kHz emission, where the sensitivity of planned detectors degrades, making a concrete observational target for dedicated kHz detectors.

We consider a progenitor modeled as a uniformly rotating supramassive NS that spins down at constant baryonic mass $M_b$ until it crosses the turning point of stability for a given sequence \cite{Friedman1988} and collapses. We employ $\alpha_0=10^{-2}$ for both models, with $\beta_0=-6$ for the lower $m_\Phi$ case, $\alpha_0=10^{-2}$ and $\beta=-10$ for the larger. The progenitors have masses of $2.296$ and $2.527$ $M_\odot$ with spin frequencies of 811 and 746 Hz, respectively. Details of the initial-data construction, the evolutionary track of the progenitor, and the numerical setup are given in the supplemental material.

\label{sec:detectability}
\textbf{Detectability of scalar radiation. }In MSTT, scalar radiation can induce two additional  polarization modes: the breathing mode $h_b$, which produces a transverse, isotropic deformation on the detector, and the longitudinal mode $h_l$, which is associated with the mass of the field and acts along the direction of propagation. These modes are not independent, since they are both sourced by the additional degree of freedom introduced by the scalar field. The strain for each mode is given by $ h_b=2\alpha_0\phi$ and  $h_l=\left(\frac{f_\Phi}{f}\right)^2 h_b$. 
 In the massless limit, $h_l$ vanishes and only the breathing mode survives, since $f_\Phi\propto m_\Phi\rightarrow 0$. 
 
The detector response to this strain depends on the location of the source on the sky-plane $(\vartheta,\varphi)$, through the angular pattern of the detector, $F(\vartheta,\varphi)$. For a breathing mode, $F_b(\vartheta,\varphi)=-\frac{1}{2}\sin^2{\vartheta}\cos{2\varphi}$ \cite{Nakao2001}, while for a longitudinal mode, $F_l(\vartheta,\varphi)=-F_b(\vartheta,\varphi)$ \cite{Nishizawa2009}. Following \cite{Rosca-Mead2020}, for a significantly long observation window, the time dependence of the sky coordinates due to the Earth's rotation can be approximated to $F_b\simeq\sqrt{4\pi/15}$.  Both modes cannot be disentangled since they only differ by a minus sign in the interferometer response.  

We propagate the scalar field extracted at a finite radius to a distance $D$ using an stationary phase approximation \cite{Rosca-Mead2020}. For an observation window $T$, such that $T\gg 1/f$, the SNR  $\rho$ for the scalar strain from a source at distance $D$ can be approximated by
\begin{equation}\label{eq:SNR}
    \rho\simeq \frac{\sqrt{T}}{\sqrt{S_n(f)}} 2\alpha_0 F_b  \frac{A(t,D)}{D} \left[1-\left(\frac{f_\Phi}{f(t)}\right)^2 \right],
\end{equation}
where $S_n(f)$ represents the instrumental noise of the detectors as a function of frequency and $A(t,D)$ is the propagated amplitude which decays with distance as $D^{-1/2}$.

\begin{figure}[t]
\includegraphics[width=0.9\linewidth]{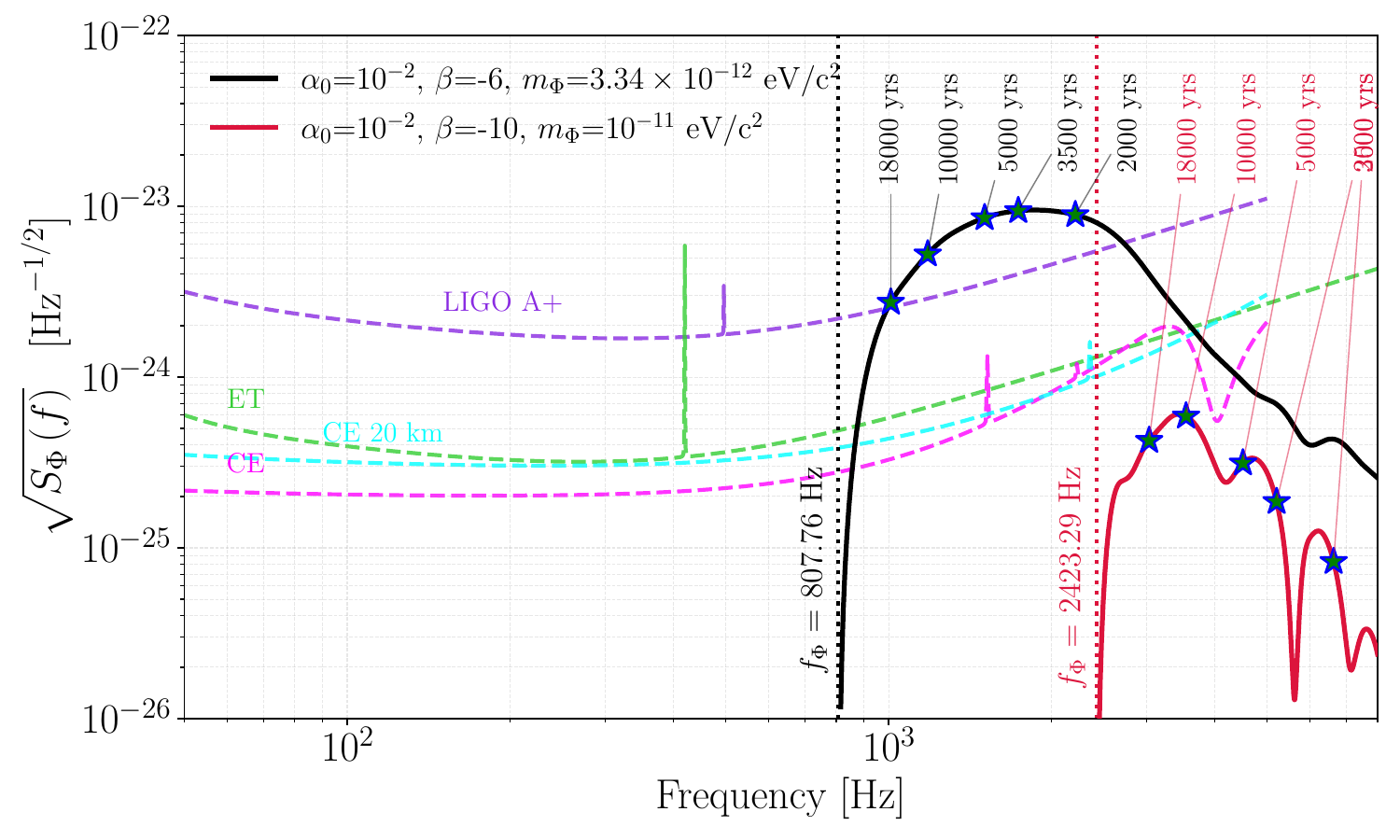}
\caption{  $\sqrt{S_\Phi}$ compared to the noise curves of A+, ET and CE detectors. The stars show the time needed for each frequency to reach the detectors for a source located at 8.5 kpc and a 10-month observation period. The black line represents the low-mass case $m_\Phi=3.3\times 10^{-12} \rm{eV/c^2}$ while the red one shows  $m_\Phi=10^{-11}\rm{eV/c^2}$. The vertical lines mark $f_\Phi$ for each case.}\label{fig:strainSF_main}
\end{figure}

Fig.~\ref{fig:strainSF_main} displays the effective spectral density $\sqrt{S_\Phi(f)}$ of the scalar signal for both scalar-field masses compared with the sensitivity curves of the Einstein Telescope (ET) \cite{ET_D_sensitivity}, LIGO's upgrade A+ \cite{Miller2015} and CE \cite{CosmicExplorer2022} (for both, the 20 and 40 km detectors). We assume ten months of observation and sources located at the Galactic Center, $D=8.5$ kpc. Since $\rho\propto D^{-3/2}\sqrt{T}$, longer observational windows would increase the accumulated SNR proportionally to $\sqrt{T}$, while considering sources closer to Earth results in an increase $\sim D^{3/2}$, contrary to the usual tensorial GWs, where the emission decays as $D^{-1}$.

Since the group velocity $v_g$ vanishes as $f\rightarrow f_\Phi$, the highest frequencies arrive first, with lower frequencies arriving progressively later as $v_g\rightarrow0$.   The first observable frequencies arrive $\sim10^{2}$~yr after the NS collapses, and the signal remains present for more than $10^{4}$~yr. However, the arrival time of the frequencies carrying the highest SNR differs strongly, for each case. The largest SNR peak of the high-mass signal reaches the detector after $10^4\,\rm{yrs}$, while for the lighter $m_\Phi$ model it takes $\sim6\times10^3\,\rm{yr}$. After the SNR peaks, the frequency decreases with time until it reaches $f_\Phi$. The SNR declines as the longitudinal mode suppresses the breathing one (since at the detector $F_l=-F_b$) when close to the threshold. Since only frequencies $f>f_\Phi$ will reach the detector, the heavier case with larger $f_\Phi$ discards a much more significant portion of the frequency content of the signal spectrum.

The light-mass model reaches $\sqrt{S_\Phi}\simeq9.8\times10^{-24}\,\mathrm{Hz^{-1/2}}$ at $\sim 1.3$ kHz, yielding a network SNR of $\sim 24$ for an ideal match-filtering search using the A+-ET-CE network. The heavy case peaks instead at $\sqrt{S_\Phi}\simeq6.3\times10^{-25}\,\mathrm{Hz^{-1/2}}$ near $3$~kHz, and lies closely below the CE curve. The post-merger tuned CE configuration will improve the detectability prospects since it is expected to have a sensitivity of $10^{-25}$ Hz$^{-1/2}$ \cite{CosmicExplorer2022}. 

The two cases differ in how much freedom remains in the parameter space. The light case sits at the bound of the region allowed by GW170817, so the amplitude reported here is already close to maximal.  However, for the heavy case, higher amplitudes than the one shown here can be achieved while still remaining consistent with current bounds, since $\sqrt{S_\Phi}\propto\alpha_0$.

\begin{figure}[t]
\includegraphics[width=0.9\linewidth]{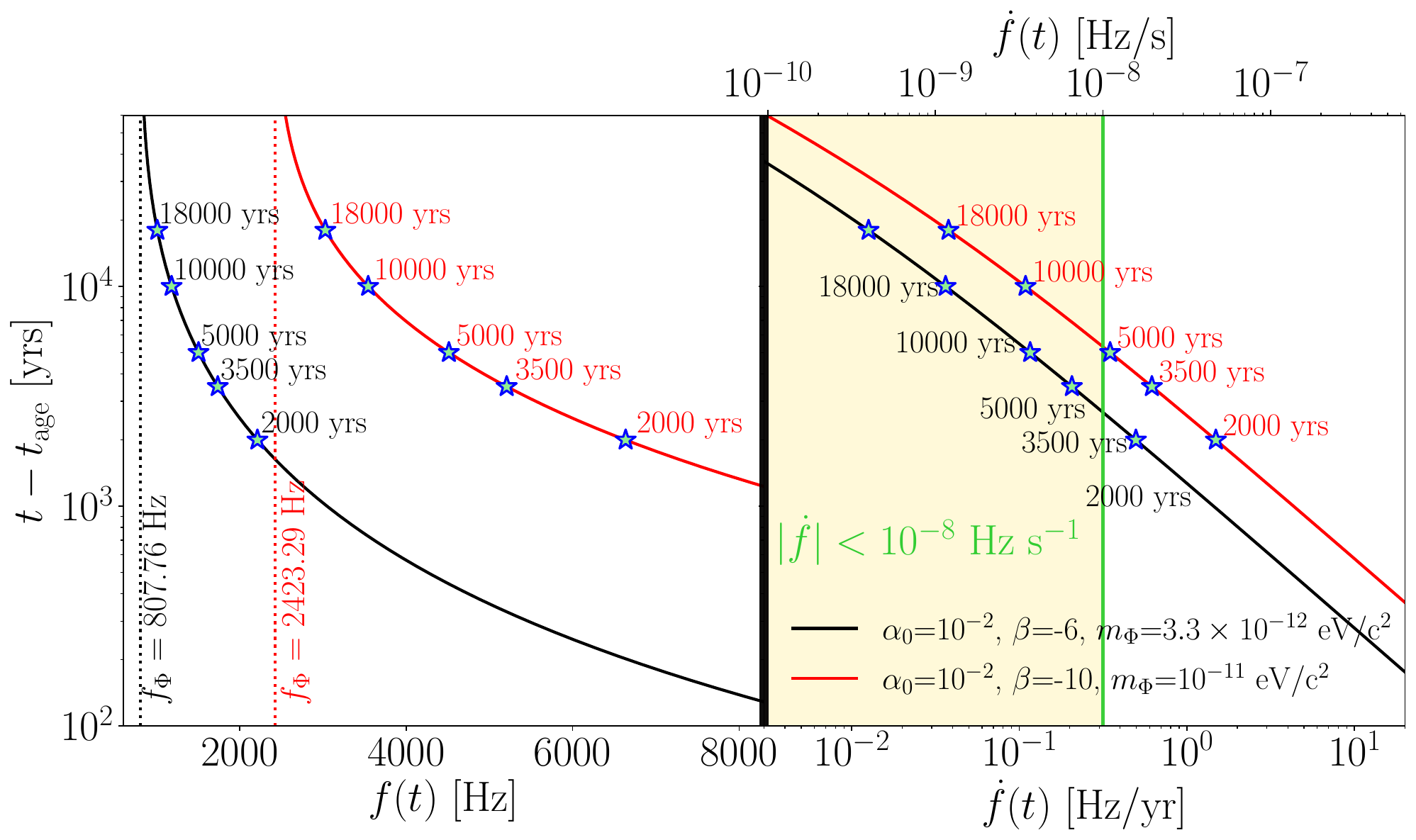}
\caption{Left: Time evolution of frequencies reaching the detector network. $t_{\rm age}$ represents the instant when the star collapses. The vertical lines represent $f_\Phi$ for each simulation. Right: Evolution of $\dot{f}(t)$ with the shaded region corresponding to $|\dot{f}|<10^{-8}\,\rm Hz\,s^{-1}$. Star symbols as in Fig.~\ref{fig:strainSF_main}.}\label{fig:fdot}
\end{figure}

Figure \ref{fig:fdot} shows (left panel) the map of each emitted frequency onto a distinct arrival time (indicated by the star symbols). The right panel shows the frequency drift for each arrival time with the golden band showing the typical values of $|\dot{f}|$ employed in CW searches \cite{LVK2022,LIGOScientific:2026plm}.  As inferred from the arrival time values, an observer at each fixed epoch will only observe a narrow slice of the emitted spectrum. When the frequency associated with the maximum value of $\sqrt{S_\Phi}$ reaches the detector, the frequency drift rate is already well below $10^{-8}\,\rm Hz\,s^{-1}$ for both simulations.

\begin{figure}
    \centering
    \includegraphics[width=0.99\linewidth]{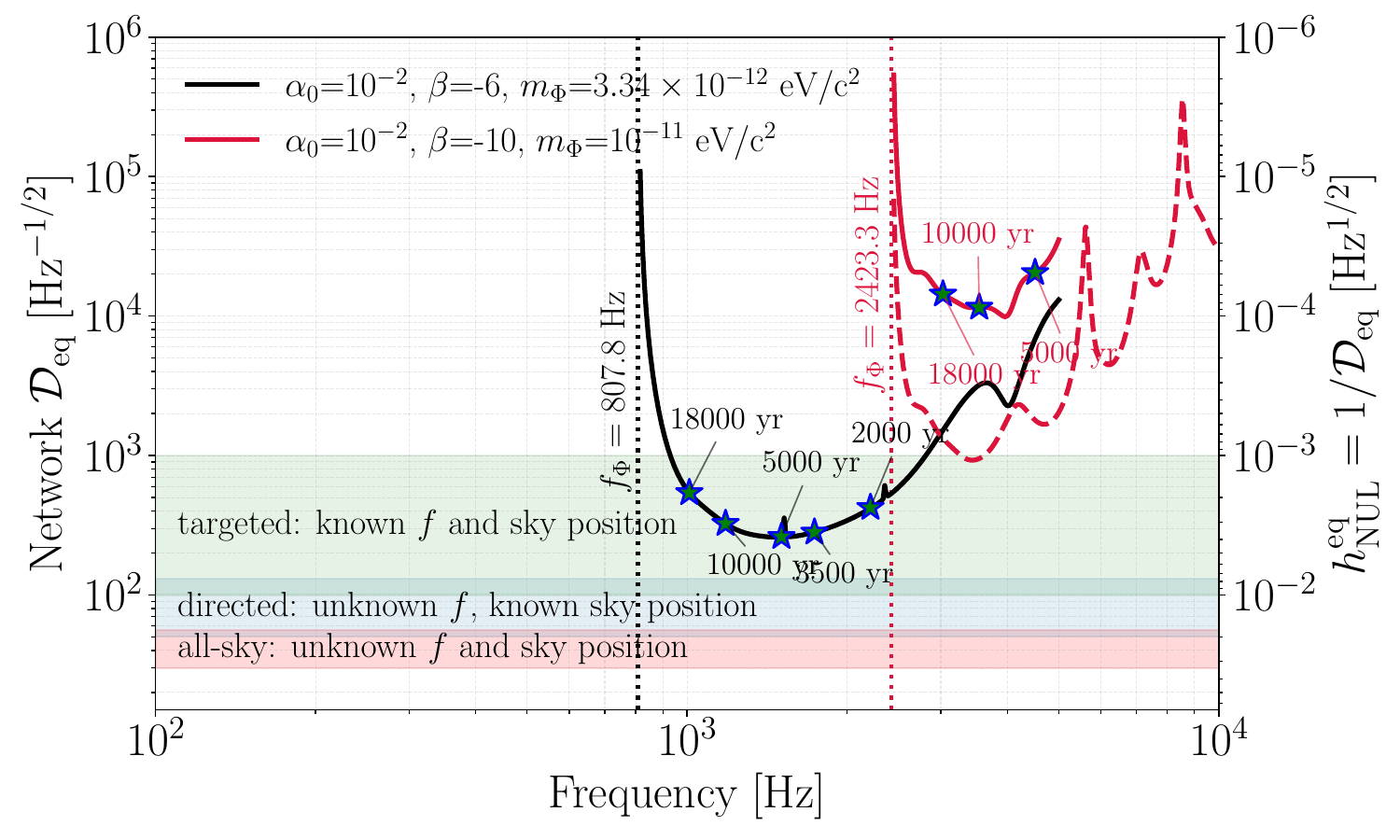}
    \caption{Sensitivity depth $\mathcal{D}_{\rm eq}(f)$ and Normalized Upper Limit $h_{\rm{NUL}}$ for the A+-ET-CE detector network.  The color bands represent the sensitivity depth achieved by LVK all-sky \cite{Steltner:2023cfk,LVK2022,LIGOScientific:2026plm}, directed \cite{Abac2026,Ming:2025ehy} and targeted CW searches \cite{Riles:2022wwz}. The vertical lines show $f_\Phi$. The red dashed line corresponds to the high-mass case assuming an idealized single detector with $\sqrt{S_n(f>2.4\rm{\,kHz)}}=10^{-25}\,\rm{Hz^{-1/2}}$.  }
    \label{fig:sebsitivity_depth}
\end{figure}

CW searches are characterized by the sensitivity depth $\mathcal{D}$ \cite{Dreissigacker:2018afk,Behnke2015}. Following its standard definition, we define the equivalent sensitivity depth needed for the scalar signal to be detected at a distance of 8.5 kpc as,
\begin{equation}\label{eq:sensitivity}
\mathcal{D}_{\rm eq}(f) = \frac{\sqrt{S_n(f)}}{h_0(f)} \simeq \frac{\sqrt{T}}{\rho(f)},
\end{equation}
 where the second equality follows from the relation $h_0(f) \simeq \rho(f) \sqrt{\frac{S_n(f)}{T}}$. The most recent searches performed by the LVK Collaboration instead use the Normalized Upper Limit (NUL) $h_{\rm{NUL}}=1/\mathcal{D}$ \cite{Abbott2021_ContinuousGW_SNR}.  For the LIGO-ET-CE network of detectors, the noise amplitude is the harmonic mean of each individual $\sqrt{S_n}$ \cite{Dreissigacker:2018afk}.

Fig.~\ref{fig:sebsitivity_depth} displays $\mathcal{D}_{\rm eq}$ calculated for the A+-ET-CE network, with colored bands representing the sensitivity depth $\mathcal{D}$ achieved in current LVK CW searches. In a CW search with fixed search parameters, the sensitivity depth is independent of the noise \cite{Dreissigacker:2018afk,Behnke2015}, so the depths achieved by LVK carry over directly to the A+-ET-CE network. Our $\mathcal{D}_{\rm eq}$ acts as a noise dependent threshold such that a search reaching $\mathcal{D}>\mathcal{D}_{\rm eq}$ would yield a detection. The O4 run reached a maximum $\mathcal{D}=93.2\,\rm{Hz^{-1/2}}$ in the search for GW from young SN remnants for frequencies around 300 Hz \cite{Abac2026}. The highest $\mathcal{D}$ achieved in a directed search is $131 \,\rm Hz^{-1/2}$ for the G347.3 source \cite{Ming:2025ehy}. Fig.~\ref{fig:sebsitivity_depth} shows that  the sensitivity depth required to detect a descalarization burst for the light-mass case should be at least $233\,\rm{Hz^{-1/2}}$, within the values needed for targeted searches using the A+-ET-CE network. The heavy case becomes marginally detectable ($\mathcal{D}_{\rm{eq}}\sim762\,\rm{Hz^{-1/2}}$) if we assume a single idealized detector ($\mathcal{D}_{\rm eq}$ decreases when considering a detector network) with  $\sqrt{S_n(f>2.4\rm{\,kHz)}}=10^{-25}\,\rm{Hz^{-1/2}}$, a highly sensitive detector in the kHz range such as NEMO \cite{Ackley:2020atn} or the kHz-tuned CE \cite{CosmicExplorer2022}.


\textbf{Observational rates and astrophysical implications. }A descalarization burst is emitted whenever a scalarized, near-maximum-mass NS collapses to a BH.
Because in our models the threshold for scalarization is well below the maximum mass, every collapsing star will be scalarized. Several channels contribute to this end state: 
the spin-down collapse of a supramassive pulsar \cite{Falcke2014,Ravi2014} (our case study), accretion-induced collapse \cite{Dermer2006}, and the delayed collapse of a cooling proto-NS during the Kelvin--Helmholtz phase \cite{Pons1999,Burrows:1986me}. Each channel is individually rare and the number of events is strongly EoS-dependent \cite{OConnor2011,Sukhbold_2016,Neustadt2021,Antoniadis:2016hxz}. Further discussion regarding each channel is shown in the Supplementary Material.

Within the $\sim 8.5\,$kpc horizon (a distance enclosing a large fraction of the Galactic stellar mass)  set by our SNR analysis for the A+-ET-CE network,  Galactic estimates for the rate in the supramassive spin-down channel span $R_{\rm{sup}} \lesssim 3\times10^{-4}\,\mathrm{yr}^{-1}$. This value is inferred from the Galactic NS birth rate and by estimating the fraction of the population that is supramassive and collapses due to spin-down. Details on this estimate are available in the supplemental material. 
Even if $R_{\rm{sup}}$ is small,  the dispersion produced by the scalar mass stretches the signal from a single event over the duration of the signal $\Delta t$. We adopt $\Delta t\sim10^{4}$~yr, based on the high-mass case. During this period, the frequency drifts slowly while remaining quasi-monochromatic. The number of detectable relic signals in transit through the Galaxy at any epoch is therefore set not only by  $R_{\rm{sup}}$ but by $ N \;\sim\; R_{\rm{sup}}\,\Delta t   \;\lesssim\; 3  $.
That is, the $\sim 10^{4}\,$yr dispersion boosts the effective source count by four orders of magnitude relative to the instantaneous event rate, so that up to $\sim$ 3 signals may be in transit across the Galaxy at any epoch. We note that our choice of \(R_{\text{sup}}\) is strictly an upper limit; order-of-magnitude uncertainties from the EoS and population estimates prevent a reliable lower bound.

The delay in the signal by $10^2-10^4$ years after NS collapse, allows to identify potential candidates. The most promising are CCSN remnants with no central pulsar. Since the observed frequency depends on time, the ideal target would be remnants of $\sim10^3-10^5$ years. X-ray surveys have identified 23 remnants without a central NS within 5 kpc~\cite{Kaplan:2004tm} and 6 within a distance of $\sim1$ kpc~\cite{Kaplan:2006mb}. Although some of these findings have been reassessed, some objects remain valid for our proposed search, e.g.~the Cygnus Loop remnant \cite{Preite2011}, with an estimated age of $2\times10^4$ years and located at $725\pm15$ pc \cite{Fesen:2021buq}, G156.2+5.7, at $\sim 1.7-3$ kpc and an age $>2\times10^4$ years \cite{Katsuda:2016ekc}, and Kes 27 at $4.3-6.5$ kpc with age $2.4\times10^{3}-10^4$ yrs \cite{Kobashi:2022vcz}.  In all cases, $\mathcal{D_{\rm eq}}$ increases with respect to our original 8.5 kpc estimated distance as $D^{3/2}$ reaching values of $\mathcal{D_{\rm eq}}<100\,\rm{Hz^{-1/2}}$. The observable frequency will depend on the age of the remnant and the time of collapse, as well as on the distance to the source and the frequency cutoff. A proper analysis of prospective sources is in preparation and will be reported elsewhere. Whether these sources can produce a supramassive NS in MSTT has been studied only for non-rotating progenitors \cite{Rosca-Mead2020}. Rotation, EoS, metallicity and the scalar couplings could change the result and are deferred for future work.

Finally, we note that unlike in pulsar searches, where the spindown parameters are phenomenological, $f(t)$ here  follows directly from the dispersion relation, so  $\dot{f}$ and $\ddot{f}$ are fixed functions of ($f,f_\Phi,D$) and, hence, $\dot{f}$ and $\ddot{f}$ are determined and fixed by ($f_\Phi,D$). For a known source, the sky location and the distance $D$ are fixed, leaving only $f_\Phi$ as a free parameter (besides the frequency $f$) and allowing longer coherent integration than usual directed searches. Since $f_\Phi$ is a fundamental constant of the theory, all emitted radiation will cluster above the same cutoff, independently of the channel and epoch of BH formation. Hence, a single detection would fix $m_\Phi$ while multiple detections would appear as signals that drift towards lower frequencies, all clustered above a common threshold, a specific signature of MSTT that cannot be reproduced by any other astrophysical population.


\textbf{Discussion and outlook.} The descalarization burst signal studied in this work is characterized by three distinct features. First, it is dispersed during its propagation, resulting in a long lasting signal ($>10^4$ years). Second, it is quasi-monochromatic, since at any given observational epoch, $|\dot{f}|< 10^{-8}\, \rm{Hz/s}$, making it an ideal candidate for CW searches. Lastly, the signal is inherently high-frequency since the frequency content is suppressed below $f_\Phi=m_\Phi c^2/h$ and even when $f_\Phi\simeq0.8$ kHz, the peak of the signal lies at $\sim1.3$ kHz. 
The detectability of the signal is not independent of observational bounds. Although the degeneracies between the scalar charge and the binary parameters may relax the GW170817 bound on the scalar field mass, conservative compatibility with observations favors $m_\Phi\gtrsim 10^{-11}\rm{eV/}c^2$, placing the bulk of the scalar signal in the $> 2.4$~kHz band where current detector sensitivity is degraded, requiring a detector optimized for the kHz band. These descalarization bursts, together with other independent signals such as the emission predicted by nuclear phase transitions in proto-NS \cite{Ecker:2026ixk} and post-merger signals \cite{LIGOScientific:2017fdd}, strengthen the case for a detector optimized for the kHz-band.

 Since scalar GWs decay as $D^{-3/2}$ because of dispersion, Galactic populations of compact objects are a natural way to test MSTT. We propose CW searches directed at Galactic SN remnants, with concrete cases such as the Cygnus Loop, G156.1+5.7 and Kes 27. As the electromagnetic counterpart of the scalar signal would arrive centuries before the scalar GW emission, catalogs could be used to identify potential candidates for directed searches. 
 The more massive the field the longer the signal's lifetime, increasing the number of signals in transit $N_{\rm sup}$. The two cases reported here lie in a mass range that is optimal for detectability while remaining compatible with currrent observations. As the mass of the field increases, the observational bounds remain satisfied, but the detectability degrades. A systematic survey over the EoS and $(\alpha_0,\beta_0,m_\Phi)$ will be presented in a companion paper and several extensions are left for future work. 

  Since $f_\Phi$ is a parameter of the theory and not of the source, any Galactic channel that involves descalarization would radiate above the same frequency cutoff and $N_{\rm sup}$ acts as a lower bound on the total number of signals in transit throughout the Galaxy. Considering CCSNe and descalarization processes in general (such as accretion-induced descalarization \cite{Kuan2022}) would increase this number. The present simulations employ a cold, uniformly rotating progenitor and neglect magnetic fields and finite-temperature effects. Extending our framework to model more realistic progenitors and SN remnants along with known pulsar spin-down estimates will facilitate targeted and injection-based CW searches. A study of the feasibility of the known SN remnants without a central pulsar will be reported elsewhere. 

 Descalarization bursts represent a new astrophysical class of persistent, quasi-monochromatic kHz radiation. Searching for them is not only feasible with current CW pipelines, but also motivates detector designs that use long-time integration to compensate for reduced high-frequency sensitivity. Whether successful or not, such searches will set robust bounds on the \((m_\Phi, \alpha_0)\) parameter space of MSTT in regimes where Solar System and binary-pulsar tests are screened.


\textbf{Acknowledgements. }Work supported by the Spanish Agencia Estatal de Investigación (grant PID2024-159689NB-C21) funded by MICIU/AEI/10.13039/501100011033 and by FEDER / EU, by the Generalitat Valenciana (Prometeo Excellence Programme grant CIPROM/2022/49), and by the European Horizon Europe staff exchange programme HORIZON-MSCA2021-SE-01 (grant NewFunFiCO-101086251). JCOM and DD acknowledge financial support via an Emmy Noether Research Group funded by the German Research
Foundation (DFG) under Grant No. DO 1771/1-1. JCOM acknowledges partial support by Secretaria de Ciencias, Humanidades y Tecnologías de México (SECIHTI). DD acknowledges financial support by the Spanish Ministry of Science and Innovation via the Ram\'on y Cajal programme (grant RYC2023-042559-I), funded by MCIN/AEI/ 10.13039/501100011033. The partial support of KP-06-N62/6 from the Bulgarian science fund is also gratefully acknowledged. The authors acknowledge support by the High Performance and Cloud Computing Group at the Zentrum für Datenverarbeitung of the University of Tübingen, the state of Baden-Württemberg through bwHPC and the German Research Foundation (DFG) through grant no. INST 37/935-1 FUGG. 
We acknowledge Discoverer PetaSC and EuroHPC JU for awarding this project access to Discoverer supercomputer resources. Calculations were also carried out on HEMUS supercomputer at the Center of Excellence in Informatics and ICT. The work was supported by the CoE of Informatics and ICT under Grant No. BG16RFPR002-1.014-0018-C01, financed by PRIDST 2021–2027 and co-financed by the European Union.

\bibliographystyle{apsrev4-2}
\bibliography{letter}

@ARTICLE{TGR1,
       author = {{The LIGO Scientific Collaboration} and {the Virgo Collaboration} and {the KAGRA Collaboration}},
        title = "{GWTC-4.0: Tests of General Relativity. I. Overview and General Tests}",
      journal = {arXiv e-prints},
         year = 2026,
        month = mar,
          eid = {arXiv:2603.19019},
        pages = {arXiv:2603.19019},
          doi = {10.48550/arXiv.2603.19019},
       eprint = {2603.19019},
archivePrefix = {arXiv},
 primaryClass = {gr-qc},
       adsurl = {https://ui.adsabs.harvard.edu/abs/2026arXiv260319019T}
}

@ARTICLE{TGR2,
       author = {{The LIGO Scientific Collaboration} and {the Virgo Collaboration} and {the KAGRA Collaboration}},
        title = "{GWTC-4.0: Tests of General Relativity. II. Parameterized Tests}",
      journal = {arXiv e-prints},
         year = 2026,
        month = mar,
          eid = {arXiv:2603.19020},
        pages = {arXiv:2603.19020},
          doi = {10.48550/arXiv.2603.19020},
archivePrefix = {arXiv},
       eprint = {2603.19020},
 primaryClass = {gr-qc},
       adsurl = {https://ui.adsabs.harvard.edu/abs/2026arXiv260319020T}
}

@ARTICLE{TGR3,
       author = {{The LIGO Scientific Collaboration} and {the Virgo Collaboration} and {the KAGRA Collaboration}},
        title = "{GWTC-4.0: Tests of General Relativity. III. Tests of the Remnants}",
      journal = {arXiv e-prints},
         year = 2026,
        month = mar,
          eid = {arXiv:2603.19021},
        pages = {arXiv:2603.19021},
          doi = {10.48550/arXiv.2603.19021},
archivePrefix = {arXiv},
       eprint = {2603.19021},
 primaryClass = {gr-qc},
       adsurl = {https://ui.adsabs.harvard.edu/abs/2026arXiv260319021T}
}

@article{LIGOScientific:2026oim,
    author = "Abac, A. G. and others",
    collaboration = "LIGO Scientific, VIRGO, KAGRA",
    title = "{GWTC-5.0: Tests of General Relativity}",
    eprint = "2607.19293",
    archivePrefix = "arXiv",
    primaryClass = "gr-qc",
    reportNumber = "LIGO-P2500781",
    month = "7",
    year = "2026",
    journal = "arxiv"
}

@article{Brans1961,
  title = {Mach's Principle and a Relativistic Theory of Gravitation},
  author = {Brans, C. and Dicke, R. H.},
  journal = {Phys. Rev.},
  volume = {124},
  issue = {3},
  pages = {925--935},
  numpages = {0},
  year = {1961},
  month = {11},
  publisher = {American Physical Society},
  doi = {10.1103/PhysRev.124.925},
  url = {https://link.aps.org/doi/10.1103/PhysRev.124.925}
}

@article{Jordan1959,
	author = {Jordan, P. },
	doi = {10.1007/BF01375155},
	id = {Jordan1959},
	journal = {Zeitschrift f{\"u}r Physik},
	number = {1},
	pages = {112--121},
	title = {Zum gegenw{\"a}rtigen Stand der Diracschen kosmologischen Hypothesen},
	url = {https://doi.org/10.1007/BF01375155},
	volume = {157},
	year = {1959}}

@article{Heisenberg2019,
   author = {Lavinia Heisenberg},
   doi = {10.1016/j.physrep.2018.11.006},
   issn = {0370-1573},
   journal = {Physics Reports},
   month = {3},
   pages = {1-113},
   publisher = {North-Holland},
   title = {A systematic approach to generalisations of General Relativity and their cosmological implications},
   volume = {796},
   url = {https://www.sciencedirect.com/science/article/pii/S0370157318303569?via%3Dihub},
   year = {2019}
}

@article{Zhao:2022vig,
    author = "Zhao, Junjie and Freire, Paulo C. C. and Kramer, Michael and Shao, Lijing and Wex, Norbert",
    title = "{Closing a spontaneous-scalarization window with binary pulsars}",
    eprint = "2201.03771",
    archivePrefix = "arXiv",
    primaryClass = "astro-ph.HE",
    doi = "10.1088/1361-6382/ac69a3",
    journal = "Class. Quant. Grav.",
    volume = "39",
    number = "11",
    pages = "11LT01",
    year = "2022"
}

@article{Freire:2024adf,
    author = "Freire, Paulo C. C. and Wex, Norbert",
    title = "{Gravity experiments with radio pulsars}",
    eprint = "2407.16540",
    archivePrefix = "arXiv",
    primaryClass = "gr-qc",
    doi = "10.1007/s41114-024-00051-y",
    journal = "Living Rev. Rel.",
    volume = "27",
    number = "1",
    pages = "5",
    year = "2024"
}

@article{Will:2014kxa,
    author = "Will, Clifford M.",
    title = "{The Confrontation between General Relativity and Experiment}",
    eprint = "1403.7377",
    archivePrefix = "arXiv",
    primaryClass = "gr-qc",
    doi = "10.12942/lrr-2014-4",
    journal = "Living Rev. Rel.",
    volume = "17",
    pages = "4",
    year = "2014"
}

@article{Bertotti:2003rm,
    author = "Bertotti, B. and Iess, L. and Tortora, P.",
    title = "{A test of general relativity using radio links with the Cassini spacecraft}",
    doi = "10.1038/nature01997",
    journal = "Nature",
    volume = "425",
    pages = "374--376",
    year = "2003"
}

@article{Kuan:2023hrh,
    author = "Kuan, Hao-Jui and Van Aelst, Karim and Lam, Alan Tsz-Lok and Shibata, Masaru",
    title = "{Binary neutron star mergers in massive scalar-tensor theory: Quasiequilibrium states and dynamical enhancement of the scalarization}",
    eprint = "2309.01709",
    archivePrefix = "arXiv",
    primaryClass = "gr-qc",
    doi = "10.1103/PhysRevD.108.064057",
    journal = "Phys. Rev. D",
    volume = "108",
    number = "6",
    pages = "064057",
    year = "2023"
}

@article{Ramazanolu2016,
   author = {Fethi M. Ramazanoğlu and Frans Pretorius},
   doi = {10.1103/PHYSREVD.93.064005/FIGURES/3/MEDIUM},
   issn = {24700029},
   issue = {6},
   journal = {Physical Review D},
   month = {3},
   pages = {064005},
   publisher = {American Physical Society},
   title = {Spontaneous scalarization with massive fields},
   volume = {93},
   url = {https://journals.aps.org/prd/abstract/10.1103/PhysRevD.93.064005},
   year = {2016}
}

@article{Yazadjiev:2016pcb,
    author = "Yazadjiev, Stoytcho S. and Doneva, Daniela D. and Popchev, Dimitar",
    title = "{Slowly rotating neutron stars in scalar-tensor theories with a massive scalar field}",
    eprint = "1602.04766",
    archivePrefix = "arXiv",
    primaryClass = "gr-qc",
    doi = "10.1103/PhysRevD.93.084038",
    journal = "Phys. Rev. D",
    volume = "93",
    number = "8",
    pages = "084038",
    year = "2016"
}

@article{Damour1993,
  title = {Nonperturbative strong-field effects in tensor-scalar theories of gravitation},
  author = {Damour, Thibault and Esposito-Far\`ese, Gilles},
  journal = {Phys. Rev. Lett.},
  volume = {70},
  issue = {15},
  pages = {2220--2223},
  numpages = {0},
  year = {1993},
  month = {Apr},
  publisher = {American Physical Society},
  doi = {10.1103/PhysRevLett.70.2220},
  url = {https://link.aps.org/doi/10.1103/PhysRevLett.70.2220}
}

@article{Unluturk:2025zie,
    author = {{\"U}nl{\"u}t{\"u}rk, K{\i}van{\c{c}} {\.I}. and Tuna, Semih and Yamak, O{\u{g}}uzhan K. and Ramazano{\u{g}}lu, Fethi M.},
    title = "{Nature of Phase Transitions and Metastability in Scalar-Tensor Theories}",
    eprint = "2502.01781",
    archivePrefix = "arXiv",
    primaryClass = "gr-qc",
    doi = "10.1103/14xz-d7xf",
    journal = "Phys. Rev. Lett.",
    volume = "135",
    number = "6",
    pages = "061401",
    year = "2025"
}

@article{Doneva2024,
  title = {Spontaneous scalarization},
  author = {Doneva, Daniela D. and Ramazano\ifmmode \breve{g}\else \u{g}\fi{}lu, Fethi M. and Silva, Hector O. and Sotiriou, Thomas P. and Yazadjiev, Stoytcho S.},
  journal = {Rev. Mod. Phys.},
  volume = {96},
  issue = {1},
  pages = {015004},
  numpages = {45},
  year = {2024},
  month = {3},
  publisher = {American Physical Society},
  doi = {10.1103/RevModPhys.96.015004},
  url = {https://link.aps.org/doi/10.1103/RevModPhys.96.015004}
}

@article{Yazadjiev:2025ezx,
    author = "Yazadjiev, Stoytcho S. and Doneva, Daniela D.",
    title = "{No-hair theorems in general relativity and scalar{\textendash}tensor theories}",
    eprint = "2505.01038",
    archivePrefix = "arXiv",
    primaryClass = "gr-qc",
    doi = "10.1142/S0218271825300046",
    journal = "Int. J. Mod. Phys. D",
    volume = "34",
    number = "09",
    pages = "2530004",
    year = "2025"
}

@article{Novak1998,
  title = {Spherical neutron star collapse toward a black hole in a tensor-scalar theory of gravity},
  author = {Novak, J\'er\^ome},
  journal = {Phys. Rev. D},
  volume = {57},
  issue = {8},
  pages = {4789--4801},
  numpages = {0},
  year = {1998},
  month = {Apr},
  publisher = {American Physical Society},
  doi = {10.1103/PhysRevD.57.4789},
  url = {https://link.aps.org/doi/10.1103/PhysRevD.57.4789}
}

@article{Kuan:2021lol,
    author = "Kuan, Hao-Jui and Doneva, Daniela D. and Yazadjiev, Stoytcho S.",
    title = "{Dynamical Formation of Scalarized Black Holes and Neutron Stars through Stellar Core Collapse}",
    eprint = "2103.11999",
    archivePrefix = "arXiv",
    primaryClass = "gr-qc",
    doi = "10.1103/PhysRevLett.127.161103",
    journal = "Phys. Rev. Lett.",
    volume = "127",
    number = "16",
    pages = "161103",
    year = "2021"
}

@article{Lam2024,
  title = {Binary neutron star mergers in massive scalar-tensor theory: Properties of postmerger remnants},
  author = {Lam, Alan Tsz-Lok and Kuan, Hao-Jui and Shibata, Masaru and Van Aelst, Karim and Kiuchi, Kenta},
  journal = {Phys. Rev. D},
  volume = {110},
  issue = {10},
  pages = {104018},
  numpages = {25},
  year = {2024},
  month = {Nov},
  publisher = {American Physical Society},
  doi = {10.1103/PhysRevD.110.104018},
  url = {https://link.aps.org/doi/10.1103/PhysRevD.110.104018}
}

@article{OlveraM:2026whf,
  title = {Rapidly rotating neutron star collapse in massive scalar-tensor theories},
  author = {Olvera M., Jos\'e Carlos and Doneva, Daniela D. and Cerd\'a-Dur\'an, Pablo and Font, Jos\'e A. and Yazadjiev, Stoytcho S.},
  journal = {Phys. Rev. D},
  volume = {114},
  issue = {2},
  pages = {024052},
  numpages = {25},
  year = {2026},
  month = {Jul},
  publisher = {American Physical Society},
  doi = {10.1103/2vtd-4559},
  url = {https://link.aps.org/doi/10.1103/2vtd-4559}
}

@article{Nishizawa2009,
  title = {Probing nontensorial polarizations of stochastic gravitational-wave backgrounds with ground-based laser interferometers},
  author = {Nishizawa, Atsushi and Taruya, Atsushi and Hayama, Kazuhiro and Kawamura, Seiji and Sakagami, Masa-aki},
  journal = {Phys. Rev. D},
  volume = {79},
  issue = {8},
  pages = {082002},
  numpages = {18},
  year = {2009},
  month = {Apr},
  publisher = {American Physical Society},
  doi = {10.1103/PhysRevD.79.082002},
  url = {https://link.aps.org/doi/10.1103/PhysRevD.79.082002}
}

@article{Rosca-Mead2020,
   author = {Roxana Rosca-Mead and Ulrich Sperhake and Christopher J. Moore and Michalis Agathos and Davide Gerosa and Christian D. Ott},
   doi = {10.1103/PhysRevD.102.044010},
   issn = {24700029},
   issue = {4},
   journal = {Physical Review D},
   month = {8},
   pages = {044010},
   publisher = {American Physical Society},
   title = {Core collapse in massive scalar-tensor gravity},
   volume = {102},
   url = {https://journals.aps.org/prd/abstract/10.1103/PhysRevD.102.044010},
   year = {2020}
}

@article{Sperhake:2017itk,
    author = "Sperhake, Ulrich and Moore, Christopher J. and Rosca, Roxana and Agathos, Michalis and Gerosa, Davide and Ott, Christian D.",
    title = "{Long-lived inverse chirp signals from core collapse in massive scalar-tensor gravity}",
    eprint = "1708.03651",
    archivePrefix = "arXiv",
    primaryClass = "gr-qc",
    doi = "10.1103/PhysRevLett.119.201103",
    journal = "Phys. Rev. Lett.",
    volume = "119",
    number = "20",
    pages = "201103",
    year = "2017"
}

@article{Cheong2019,
  title = {Numerical studies on core collapse supernova in self-interacting massive scalar-tensor gravity},
  author = {Cheong, Patrick Chi-Kit and Li, Tjonnie Guang Feng},
  journal = {Phys. Rev. D},
  volume = {100},
  issue = {2},
  pages = {024027},
  numpages = {6},
  year = {2019},
  month = {Jul},
  publisher = {American Physical Society},
  doi = {10.1103/PhysRevD.100.024027},
  url = {https://link.aps.org/doi/10.1103/PhysRevD.100.024027}
}

@article{Huang2021,
   doi = {10.1088/1361-6382/ac35ab},
   issn = {13616382},
   issue = {24},
   journal = {Classical and Quantum Gravity},
   month = {11},
   pages = {245006},
   publisher = {IOP Publishing},
   author = "Huang, Da and Geng, Chao-Qiang and Kuan, Hao-Jui",
   title = {Scalar gravitational wave signals from core collapse in massive scalar–tensor gravity with triple-scalar interactions},
   volume = {38},
   url = {https://iopscience.iop.org/article/10.1088/1361-6382/ac35ab https://iopscience.iop.org/article/10.1088/1361-6382/ac35ab/meta},
   year = {2021}
}

@article{Asakawa:2023obq,
    author = "Asakawa, Naomichi and Sekiguchi, Yuichiro",
    title = "{Constraining self-interactions of a massive scalar field using scalar gravitational waves from stellar core collapse}",
    eprint = "2308.15052",
    archivePrefix = "arXiv",
    primaryClass = "gr-qc",
    doi = "10.1103/PhysRevD.108.044060",
    journal = "Phys. Rev. D",
    volume = "108",
    number = "4",
    pages = "044060",
    year = "2023"
}

@article{Geng:2020slq,
    author = "Geng, Chao-Qiang and Kuan, Hao-Jui and Luo, Ling-Wei",
    title = "{Inverse-chirp imprint of gravitational wave signals in scalar tensor theory}",
    eprint = "2005.11629",
    archivePrefix = "arXiv",
    primaryClass = "gr-qc",
    doi = "10.1140/epjc/s10052-020-8359-y",
    journal = "Eur. Phys. J. C",
    volume = "80",
    number = "8",
    pages = "780",
    year = "2020"
}

@article{Rosca-Mead:2019seq,
    author = "Rosca-Mead, Roxana and Moore, Christopher J. and Agathos, Michalis and Sperhake, Ulrich",
    title = "{Inverse-chirp signals and spontaneous scalarisation with self-interacting potentials in stellar collapse}",
    eprint = "1903.09704",
    archivePrefix = "arXiv",
    primaryClass = "gr-qc",
    doi = "10.1088/1361-6382/ab256f",
    journal = "Class. Quant. Grav.",
    volume = "36",
    number = "13",
    pages = "134003",
    year = "2019"
}

@article{LIGOScientific:2017fdd,
    author = "Abbott, B. P. and others",
    collaboration = "LIGO Scientific, Virgo",
    title = "{Search for Post-merger Gravitational Waves from the Remnant of the Binary Neutron Star Merger GW170817}",
    eprint = "1710.09320",
    archivePrefix = "arXiv",
    primaryClass = "astro-ph.HE",
    reportNumber = "LIGO-P1700318",
    doi = "10.3847/2041-8213/aa9a35",
    journal = "Astrophys. J. Lett.",
    volume = "851",
    number = "1",
    pages = "L16",
    year = "2017"
}

@article{ET:2025xjr,
    author = "Abac, Adrian and others",
    collaboration = "ET",
    title = "{The Science of the Einstein Telescope}",
    eprint = "2503.12263",
    archivePrefix = "arXiv",
    primaryClass = "gr-qc",
    reportNumber = "ET-0036C-25",
    doi = "10.1088/1475-7516/2026/03/081",
    journal = "JCAP",
    volume = "03",
    pages = "081",
    year = "2026"
}

@article{LISA:2022kgy,
    author = "Arun, K. G. and others",
    collaboration = "LISA",
    title = "{New horizons for fundamental physics with LISA}",
    eprint = "2205.01597",
    archivePrefix = "arXiv",
    primaryClass = "gr-qc",
    doi = "10.1007/s41114-022-00036-9",
    journal = "Living Rev. Rel.",
    volume = "25",
    number = "1",
    pages = "4",
    year = "2022"
}

@article{ET_D_sensitivity,
   doi = {10.1088/0264-9381/28/9/094013},
   issn = {0264-9381},
   issue = {9},
   author = "Hild, S. and others",
   journal = {Classical and Quantum Gravity},
   month = {4},
   pages = {094013},
   publisher = {IOP Publishing},
   title = {Sensitivity studies for third-generation gravitational wave observatories},
   volume = {28},
   url = {https://iopscience.iop.org/article/10.1088/0264-9381/28/9/094013 https://iopscience.iop.org/article/10.1088/0264-9381/28/9/094013/meta},
   year = {2011}
}

@article{CosmicExplorer2022,
   author = {Varun Srivastava and Derek Davis and Kevin Kuns and Philippe Landry and Stefan Ballmer and Matthew Evans and Evan D. Hall and Jocelyn Read and B. S. Sathyaprakash},
   doi = {10.3847/1538-4357/AC5F04},
   issn = {0004-637X},
   issue = {1},
   journal = {The Astrophysical Journal},
   month = {5},
   pages = {22},
   publisher = {IOP Publishing},
   title = {Science-driven Tunable Design of Cosmic Explorer Detectors},
   volume = {931},
   url = {https://iopscience.iop.org/article/10.3847/1538-4357/ac5f04 https://iopscience.iop.org/article/10.3847/1538-4357/ac5f04/meta},
   year = {2022}
}

@ARTICLE{Aggarwal:2025,
       author = {{Aggarwal}, Nancy and {Aguiar}, Odylio D. and {Blas}, Diego and {Bauswein}, Andreas and {Cella}, Giancarlo and {Clesse}, Sebastian and {Cruise}, Adrian Michael and {Domcke}, Valerie and {Ellis}, Sebastian and {Figueroa}, Daniel G. and {Franciolini}, Gabriele and {Garc{\'\i}a-Cely}, Camilo and {Geraci}, Andrew and {Goryachev}, Maxim and {Grote}, Hartmut and {Hindmarsh}, Mark and {Ito}, Asuka and {Kopp}, Joachim and {Lee}, Sung Mook and {Martineau}, Killian and {McDonald}, Jamie and {Muia}, Francesco and {Mukund}, Nikhil and {Ottaway}, David and {Peloso}, Marco and {Peters}, Krisztian and {Quevedo}, Fernando and {Ricciardone}, Angelo and {Ringwald}, Andreas and {Steinlechner}, Jessica and {Steinlechner}, Sebastian and {Sun}, Sichun and {Tamarit}, Carlos and {Tobar}, Michael E. and {Torrenti}, Francisco and {{\"U}nal}, Caner and {White}, Graham},
        title = "{Challenges and opportunities of gravitational-wave searches above 10 kHz}",
      journal = {Living Reviews in Relativity},
         year = 2025,
        month = nov,
       volume = {28},
       number = {1},
          eid = {10},
        pages = {10},
          doi = {10.1007/s41114-025-00060-5},
archivePrefix = {arXiv},
       eprint = {2501.11723},
 primaryClass = {gr-qc},
       adsurl = {https://ui.adsabs.harvard.edu/abs/2025LRR....28...10A}
}

@article{Wette:2023dom,
    author = "Wette, Karl",
    title = "{Searches for continuous gravitational waves from neutron stars: A twenty-year retrospective}",
    eprint = "2305.07106",
    archivePrefix = "arXiv",
    primaryClass = "gr-qc",
    reportNumber = "LIGO-P2300131-v4, LIGO P2300131-v3",
    doi = "10.1016/j.astropartphys.2026.103229",
    journal = "Astropart. Phys.",
    volume = "153",
    pages = "102880",
    year = "2023",
    note = "[Erratum: Astropart.Phys. 179, 103229 (2026)]"
}

@article{Riles:2022wwz,
    author = "Riles, Keith",
    title = "{Searches for continuous-wave gravitational radiation}",
    eprint = "2206.06447",
    archivePrefix = "arXiv",
    primaryClass = "astro-ph.HE",
    doi = "10.1007/s41114-023-00044-3",
    journal = "Living Rev. Rel.",
    volume = "26",
    number = "1",
    pages = "3",
    year = "2023"
}

@article{Dreissigacker:2018afk,
    author = "Dreissigacker, Christoph and Prix, Reinhard and Wette, Karl",
    title = "{Fast and Accurate Sensitivity Estimation for Continuous-Gravitational-Wave Searches}",
    eprint = "1808.02459",
    archivePrefix = "arXiv",
    primaryClass = "gr-qc",
    doi = "10.1103/PhysRevD.98.084058",
    journal = "Phys. Rev. D",
    volume = "98",
    number = "8",
    pages = "084058",
    year = "2018"
}

@article{Behnke2015,
  title = {Postprocessing methods used in the search for continuous gravitational-wave signals from the Galactic Center},
  author = {Behnke, Berit and Papa, Maria Alessandra and Prix, Reinhard},
  journal = {Phys. Rev. D},
  volume = {91},
  issue = {6},
  pages = {064007},
  numpages = {17},
  year = {2015},
  month = {Mar},
  publisher = {American Physical Society},
  doi = {10.1103/PhysRevD.91.064007},
  url = {https://link.aps.org/doi/10.1103/PhysRevD.91.064007}
}

@article{LVK2022,
  title = {All-sky search for continuous gravitational waves from isolated neutron stars using Advanced LIGO and Advanced Virgo O3 data},
  author = {Abbott, R. and others},
  collaboration = {LIGO Scientific Collaboration, Virgo Collaboration, and KAGRA Collaboration},
  journal = {Phys. Rev. D},
  volume = {106},
  issue = {10},
  pages = {102008},
  numpages = {37},
  year = {2022},
  month = {Nov},
  publisher = {American Physical Society},
  doi = {10.1103/PhysRevD.106.102008},
  url = {https://link.aps.org/doi/10.1103/PhysRevD.106.102008}
}

@article{LIGOScientific:2026plm,
    author = "Abac, A. G. and others",
    collaboration = "LIGO Scientific, VIRGO, KAGRA",
    title = "{All-sky Searches for Continuous Gravitational Waves from Isolated Neutron Stars in the Data from the First Part of the Fourth LIGO-Virgo-KAGRA Observing Run}",
    eprint = "2603.14168",
    journal="arxiv",
    archivePrefix = "arXiv",
    primaryClass = "gr-qc",
    reportNumber = "LIGO-P2500416",
    month = "3",
    year = "2026"
}

@article{Steltner:2023cfk,
    author = "Steltner, B. and Papa, M. A. and Eggenstein, H. -B. and Prix, R. and Bensch, M. and Allen, B. and Machenschalk, B.",
    title = "{Deep Einstein@Home All-sky Search for Continuous Gravitational Waves in LIGO O3 Public Data}",
    eprint = "2303.04109",
    archivePrefix = "arXiv",
    primaryClass = "gr-qc",
    doi = "10.3847/1538-4357/acdad4",
    journal = "Astrophys. J.",
    volume = "952",
    number = "1",
    pages = "55",
    year = "2023"
}

@article{Romani2022,
   author = {Roger W. Romani and D. Kandel and Alexei V. Filippenko and Thomas G. Brink and WeiKang Zheng},
   doi = {10.3847/2041-8213/AC8007},
   issn = {2041-8205},
   issue = {2},
   journal = {The Astrophysical Journal Letters},
   month = {7},
   pages = {L17},
   publisher = {IOP Publishing},
   title = {PSR J0952−0607: The Fastest and Heaviest Known Galactic Neutron Star},
   volume = {934},
   url = {https://iopscience.iop.org/article/10.3847/2041-8213/ac8007 https://iopscience.iop.org/article/10.3847/2041-8213/ac8007/meta},
   year = {2022}
}

@article{Romani2025,
   author = {Roger W. Romani and Maya Beleznay and Alexei V. Filippenko and Thomas G. Brink and WeiKang Zheng},
   doi = {10.3847/1538-4357/AE28C5},
   issn = {0004-637X},
   issue = {1},
   journal = {The Astrophysical Journal},
   month = {12},
   pages = {101},
   publisher = {IOP Publishing},
   title = {PSR J0952-0607: Tightening a Record-high Neutron Star Mass},
   volume = {996},
   url = {https://iopscience.iop.org/article/10.3847/1538-4357/ae28c5 https://iopscience.iop.org/article/10.3847/1538-4357/ae28c5/meta},
   year = {2025}
}

@article{Doneva2013,
  title = {Rapidly rotating neutron stars in scalar-tensor theories of gravity},
  author = {Doneva, Daniela D. and Yazadjiev, Stoytcho S. and Stergioulas, Nikolaos and Kokkotas, Kostas D.},
  journal = {Phys. Rev. D},
  volume = {88},
  issue = {8},
  pages = {084060},
  numpages = {12},
  year = {2013},
  month = {10},
  publisher = {American Physical Society},
  doi = {10.1103/PhysRevD.88.084060},
  url = {https://link.aps.org/doi/10.1103/PhysRevD.88.084060}
}

@article{Abac2026,
   author = {A. G. Abac and others (The LIGO Scientific Collaboration},
   doi = {10.3847/1538-4357/AE6263},
   issn = {0004-637X},
   issue = {2},
   journal = {The Astrophysical Journal},
   month = {6},
   pages = {204},
   publisher = {IOP Publishing},
   title = {Searches for Continuous Gravitational Waves from Supernova Remnants in the First Part of the LIGO-Virgo-KAGRA Fourth Observing Run},
   volume = {1004},
   url = {https://iopscience.iop.org/article/10.3847/1538-4357/ae6263 https://iopscience.iop.org/article/10.3847/1538-4357/ae6263/meta},
   year = {2026}
}

@article{Ming:2025ehy,
    author = "Ming, Jing and Papa, Maria Alessandra and Eggenstein, Heinz-Bernd and Machenschalk, Bernd and Martins, Jasper and Steltner, Benjamin and McGloughlin, Brian and Dergachev, Vladimir and Prix, Reinhard and Bensch, Maximillian",
    title = "{Observational Constraints on the Spin/Anisotropy of the Central Compact Objects of Cassiopeia A, Vela Jr., and G347.3{\textendash}0.5 and a Single Surviving Continuous Gravitational Wave Candidate}",
    eprint = "2512.15672",
    archivePrefix = "arXiv",
    primaryClass = "gr-qc",
    doi = "10.3847/1538-4357/ae6e41",
    journal = "Astrophys. J.",
    volume = "1004",
    number = "1",
    pages = "65",
    year = "2026"
}

@article{Abbott2021_ContinuousGW_SNR,
   author = {R. Abbott and others},
   doi = {10.3847/1538-4357/AC17EA},
   issn = {0004-637X},
   issue = {1},
   journal = {The Astrophysical Journal},
   month = {11},
   pages = {80},
   publisher = {IOP Publishing},
   title = {Searches for Continuous Gravitational Waves from Young Supernova Remnants in the Early Third Observing Run of Advanced LIGO and Virgo},
   volume = {921},
   url = {https://iopscience.iop.org/article/10.3847/1538-4357/ac17ea https://iopscience.iop.org/article/10.3847/1538-4357/ac17ea/meta},
   year = {2021}
}

@article{Damour1992,
   author = {T. Damour and G. Esposito-Farese},
   doi = {10.1088/0264-9381/9/9/015},
   issn = {0264-9381},
   issue = {9},
   journal = {Classical and Quantum Gravity},
   month = {9},
   pages = {2093},
   publisher = {IOP Publishing},
   title = {Tensor-multi-scalar theories of gravitation},
   volume = {9},
   url = {https://iopscience.iop.org/article/10.1088/0264-9381/9/9/015 https://iopscience.iop.org/article/10.1088/0264-9381/9/9/015/meta},
   year = {1992}
}

@article{Batrakov2024,
   author = {A. Batrakov and H. Hu and N. Wex and P. C.C. Freire and V. Venkatraman Krishnan and M. Kramer and Y. J. Guo and L. Guillemot and J. W. McKee and I. Cognard and G. Theureau},
   doi = {10.1051/0004-6361/202245246},
   issn = {0004-6361},
   journal = {Astronomy \& Astrophysics},
   month = {6},
   pages = {A101},
   publisher = {EDP Sciences},
   title = {A new pulsar timing model for scalar-tensor gravity with applications to PSR J2222-0137 and pulsar-black hole binaries},
   volume = {686},
   url = {https://www.aanda.org/articles/aa/full_html/2024/06/aa45246-22/aa45246-22.html https://www.aanda.org/articles/aa/abs/2024/06/aa45246-22/aa45246-22.html},
   year = {2024}
}

@article{Damour1996,
  title = {Tensor-scalar gravity and binary-pulsar experiments},
  author = {Damour, Thibault and Esposito-Far\`ese, Gilles},
  journal = {Phys. Rev. D},
  volume = {54},
  issue = {2},
  pages = {1474--1491},
  numpages = {0},
  year = {1996},
  month = {Jul},
  publisher = {American Physical Society},
  doi = {10.1103/PhysRevD.54.1474},
  url = {https://link.aps.org/doi/10.1103/PhysRevD.54.1474}
}

@article{Sanger2026,
  title = {Tests of general relativity with GW230529: A neutron star merging with a lower mass-gap compact object},
  author = {S\"anger, Elise M. and Roy, Soumen and Agathos, Michalis and Birnholtz, Ofek and Buonanno, Alessandra and Dietrich, Tim and Haney, Maria and Juli\'e, F\'elix-Louis and Pratten, Geraint and Steinhoff, Jan and Van Den Broeck, Chris and Biscoveanu, Sylvia and Char, Prasanta and Heffernan, Anna and Joshi, Prathamesh and Kedia, Atul and Schofield, R. M. S. and Trevor, M. and Zevin, Michael},
  journal = {Phys. Rev. D},
  volume = {113},
  issue = {8},
  pages = {084070},
  numpages = {20},
  year = {2026},
  month = {Apr},
  publisher = {American Physical Society},
  doi = {10.1103/r43k-51yq},
  url = {https://link.aps.org/doi/10.1103/r43k-51yq}
}

@article{Niu:2021nic,
    author = "Niu, Rui and Zhang, Xing and Wang, Bo and Zhao, Wen",
    title = "{Constraining Scalar-tensor Theories Using Neutron Star{\textendash}Black Hole Gravitational Wave Events}",
    eprint = "2105.13644",
    archivePrefix = "arXiv",
    primaryClass = "gr-qc",
    doi = "10.3847/1538-4357/ac1d4f",
    journal = "Astrophys. J.",
    volume = "921",
    number = "2",
    pages = "149",
    year = "2021"
}

@article{Alsing:2011er,
    author = "Alsing, Justin and Berti, Emanuele and Will, Clifford M. and Zaglauer, Helmut",
    title = "{Gravitational radiation from compact binary systems in the massive Brans-Dicke theory of gravity}",
    eprint = "1112.4903",
    archivePrefix = "arXiv",
    primaryClass = "gr-qc",
    doi = "10.1103/PhysRevD.85.064041",
    journal = "Phys. Rev. D",
    volume = "85",
    pages = "064041",
    year = "2012"
}

@article{LIGO2019_GW170817,
  title = {Properties of the Binary Neutron Star Merger GW170817},
  author = {Abbott, B. P. and others},
  collaboration = {LIGO Scientific Collaboration and Virgo Collaboration},
  journal = {Phys. Rev. X},
  volume = {9},
  issue = {1},
  pages = {011001},
  numpages = {32},
  year = {2019},
  month = {Jan},
  publisher = {American Physical Society},
  doi = {10.1103/PhysRevX.9.011001},
  url = {https://link.aps.org/doi/10.1103/PhysRevX.9.011001}
}

@article{Friedman1988,
   author = {John L. Friedman and James R. Ipser and Rafael D. Sorkin },
   doi = {10.1086/166043},
   issn = {0004-637X},
   journal = {ApJ},
   month = {2},
   pages = {722},
   publisher = {American Astronomical Society},
   title = {Turning Point Method for Axisymmetric Stability of Rotating Relativistic Stars},
   volume = {325},
   url = {https://ui.adsabs.harvard.edu/abs/1988ApJ...325..722F/abstract},
   year = {1988}
}

@article{Nakao2001,
  title = {Response of interferometric detectors to scalar gravitational waves},
  author = {Nakao, Ken-ichi and Harada, Tomohiro and Shibata, Masaru and Kawamura, Seiji and Nakamura, Takashi},
  journal = {Phys. Rev. D},
  volume = {63},
  issue = {8},
  pages = {082001},
  numpages = {14},
  year = {2001},
  month = {Mar},
  publisher = {American Physical Society},
  doi = {10.1103/PhysRevD.63.082001},
  url = {https://link.aps.org/doi/10.1103/PhysRevD.63.082001}
}

@article{Miller2015,
  author  = {Miller, J. and Barsotti, L. and Vitale, S. and Fritschel, P.
             and Evans, M. and Sigg, D.},
  title   = {Prospects for doubling the range of {A}dvanced {LIGO}},
  journal = {Phys. Rev. D},
  volume  = {91},
  pages   = {062005},
  year    = {2015},
  doi     = {10.1103/PhysRevD.91.062005},
  eprint  = {1410.5882},
  archivePrefix = {arXiv},
  primaryClass  = {astro-ph.IM}
}

@article{Ackley:2020atn,
    author = "Ackley, K. and others",
    title = "{Neutron Star Extreme Matter Observatory: A kilohertz-band gravitational-wave detector in the global network}",
    eprint = "2007.03128",
    archivePrefix = "arXiv",
    primaryClass = "astro-ph.HE",
    doi = "10.1017/pasa.2020.39",
    journal = "Publ. Astron. Soc. Austral.",
    volume = "37",
    pages = "e047",
    year = "2020"
}

@article{Ravi2014,
   author = {Vikram Ravi and Paul D. Lasky},
   doi = {10.1093/MNRAS/STU720},
   issn = {0035-8711},
   issue = {3},
   journal = {Monthly Notices of the Royal Astronomical Society},
   month = {7},
   pages = {2433-2439},
   publisher = {Oxford Academic},
   title = {The birth of black holes: neutron star collapse times, gamma-ray bursts and fast radio bursts},
   volume = {441},
   url = {https://dx.doi.org/10.1093/mnras/stu720},
   year = {2014}
}

@article{Falcke2014,
   author = {Heino Falcke and Luciano Rezzolla},
   doi = {10.1051/0004-6361/201321996},
   issn = {0004-6361},
   journal = {Astronomy \& Astrophysics},
   month = {2},
   pages = {A137},
   publisher = {EDP Sciences},
   title = {Fast radio bursts: the last sign of supramassive neutron stars},
   volume = {562},
   url = {https://www.aanda.org/articles/aa/full_html/2014/02/aa21996-13/aa21996-13.html https://www.aanda.org/articles/aa/abs/2014/02/aa21996-13/aa21996-13.html},
   year = {2014}
}

@article{Dermer2006,
   author = {Charles D Dermer and Armen Atoyan},
   journal = {The Astrophysical Journal},
   pages = {13-16},
   title = {COLLAPSE OF NEUTRON STARS TO BLACK HOLES IN BINARY SYSTEMS: A MODEL FOR SHORT GAMMA-RAY BURSTS},
   volume = {643},
   year = {2006}
}

@article{Pons1999,
   author = {J. A. Pons and S. Reddy and M. Prakash and J. M. Lattimer and J. A. Miralles},
   doi = {10.1086/306889},
   issn = {0004-637X},
   issue = {2},
   journal = {The Astrophysical Journal},
   month = {3},
   pages = {780},
   publisher = {IOP Publishing},
   title = {Evolution of Proto-Neutron Stars},
   volume = {513},
   url = {https://iopscience.iop.org/article/10.1086/306889 https://iopscience.iop.org/article/10.1086/306889/meta},
   year = {1999}
}

@article{Burrows:1986me,
    author = "Burrows, Adam and Lattimer, James M.",
    title = "{The birth of neutron stars}",
    doi = "10.1086/164405",
    journal = "Astrophys. J.",
    volume = "307",
    pages = "178--196",
    year = "1986"
}

@article{OConnor2011,
   author = {Evan O'Connor and Christian D. Ott},
   doi = {10.1088/0004-637X/730/2/70},
   issn = {0004-637X},
   issue = {2},
   journal = {The Astrophysical Journal},
   month = {3},
   pages = {70},
   publisher = {IOP Publishing},
   title = {BLACK HOLE FORMATION IN FAILING CORE-COLLAPSE SUPERNOVAE},
   volume = {730},
   url = {https://iopscience.iop.org/article/10.1088/0004-637X/730/2/70 https://iopscience.iop.org/article/10.1088/0004-637X/730/2/70/meta},
   year = {2011}
}

@article{Neustadt2021,
    author = {Neustadt, J M M and Kochanek, C S and Stanek, K Z and Basinger, C and Jayasinghe, T and Garling, C T and Adams, S M and Gerke, J},
    title = {The search for failed supernovae with the Large Binocular Telescope: a new candidate and the failed SN fraction with 11 yr of data},
    journal = {Monthly Notices of the Royal Astronomical Society},
    volume = {508},
    number = {1},
    pages = {516-528},
    year = {2021},
    month = {11},
    issn = {0035-8711},
    doi = {10.1093/mnras/stab2605},
    url = {https://doi.org/10.1093/mnras/stab2605},
    eprint = {https://academic.oup.com/mnras/article-pdf/508/1/516/40440715/stab2605.pdf},
}

@article{Sukhbold_2016,
	author = {Sukhbold, Tuguldur and Ertl, T. and Woosley, S. E. and Brown, Justin M. and Janka, H.-T.},
	doi = {10.3847/0004-637X/821/1/38},
	journal = {The Astrophysical Journal},
	month = {apr},
	number = {1},
	pages = {38},
	publisher = {The American Astronomical Society},
	title = {CORE-COLLAPSE SUPERNOVAE FROM 9 TO 120 SOLAR MASSES BASED ON NEUTRINO-POWERED EXPLOSIONS},
	url = {https://doi.org/10.3847/0004-637X/821/1/38},
	volume = {821},
	year = {2016}}

@article{Antoniadis:2016hxz,
    author = "Antoniadis, John and Tauris, Thomas M. and Ozel, Feryal and Barr, Ewan and Champion, David J. and Freire, Paulo C. C.",
    title = "{The millisecond pulsar mass distribution: Evidence for bimodality and constraints on the maximum neutron star mass}",
    eprint = "1605.01665",
    archivePrefix = "arXiv",
    primaryClass = "astro-ph.HE",
    month = "5",
    year = "2016",
    journal = {arXiv e-prints},
}

@article{Kaplan:2006mb,
    author = "Kaplan, David L. and Gaensler, B. M. and Kulkarni, S. R. and Slane, P. O.",
    title = "{An x-ray search for compact central sources in supernova remnants II: six large diameter snrs}",
    eprint = "astro-ph/0602312",
    archivePrefix = "arXiv",
    doi = "10.1086/501441",
    journal = "Astrophys. J. Suppl.",
    volume = "163",
    pages = "344",
    year = "2006"
}

@article{Kaplan:2004tm,
    author = "Kaplan, David L. and Frail, D. A. and Gaensler, B. M. and Gotthelf, E. V. and Kulkarni, S. R. and Slane, P. O. and Nechita, A.",
    title = "{An X-ray search for compact central sources in supernova remnants 1: SNRs G093.3+6.9, G315.4-2.3, G084.2+0.8, {\&} G127.1+0.5}",
    eprint = "astro-ph/0403313",
    archivePrefix = "arXiv",
    doi = "10.1086/421065",
    journal = "Astrophys. J. Suppl.",
    volume = "153",
    pages = "269",
    year = "2004"
}

@article{Fesen:2021buq,
    author = "Fesen, Robert A. and Weil, Kathryn E. and Cisneros, Ignacio and Blair, William P. and Raymond, John C.",
    title = "{An updated distance to the Cygnus Loop based on Gaia Early DR3}",
    eprint = "2109.05368",
    archivePrefix = "arXiv",
    primaryClass = "astro-ph.HE",
    doi = "10.1093/mnras/stab2066",
    journal = "Mon. Not. Roy. Astron. Soc.",
    volume = "507",
    number = "1",
    pages = "244--245",
    year = "2021"
}

@ARTICLE{Preite2011,
       author = {{Preite Martinez}, A.},
        title = "{The Cygnus Loop: a weak core-collapse SN in our Galaxy}",
      journal = {Astronomy \& Astrophysics},
         year = 2011,
        month = mar,
       volume = {527},
          eid = {A55},
        pages = {A55},
          doi = {10.1051/0004-6361/201015213},
       adsurl = {https://ui.adsabs.harvard.edu/abs/2011A&A...527A..55P}
}

@article{Katsuda:2016ekc,
    author = "Katsuda, Satoru and Tanaka, Masaomi and Morokuma, Tomoki and Fesen, Robert and Milisavljevic, Dan",
    title = "{Constraining the Age and Distance of the Galactic Supernova Remnant G156.2+5.7 by H$\alpha$ Expansion Measurements}",
    eprint = "1605.07271",
    archivePrefix = "arXiv",
    primaryClass = "astro-ph.HE",
    doi = "10.3847/0004-637X/826/2/108",
    journal = "Astrophys. J.",
    volume = "826",
    number = "2",
    pages = "108",
    year = "2016"
}

@article{Kobashi:2022vcz,
    author = "Kobashi, Ryosuke and Yasuda, Haruo and Lee, Shiu-Hang",
    title = "{Long-term Evolution of Nonthermal Emission from Type Ia and Core-collapse Supernova Remnants in a Diversified Circumstellar Medium}",
    eprint = "2207.06203",
    archivePrefix = "arXiv",
    primaryClass = "astro-ph.HE",
    doi = "10.3847/1538-4357/ac80f9",
    journal = "Astrophys. J.",
    volume = "936",
    number = "1",
    pages = "26",
    year = "2022"
}

@article{Ecker:2026ixk,
    author = "Ecker, Christian and Giliberti, Mauro and Rezzolla, Luciano",
    title = "{Multi-messenger and multi-band signal from first-order phase transitions in proto-neutron stars}",
    eprint = "2609.29761",
    archivePrefix = "arXiv",
    primaryClass = "gr-qc",
    month = "9",
    year = "2026",
    journal = {arxiv}
}

@article{Kuan2022,
   author = {Hao Jui Kuan and Arthur G. Suvorov and Daniela D. Doneva and Stoytcho S. Yazadjiev},
   doi = {10.1103/PhysRevLett.129.121104},
   issn = {10797114},
   issue = {12},
   journal = {Physical Review Letters},
   month = {9},
   pages = {121104},
   pmid = {36179164},
   publisher = {American Physical Society},
   title = {Gravitational Waves from Accretion-Induced Descalarization in Massive Scalar-Tensor Theory},
   volume = {129},
   url = {https://journals.aps.org/prl/abstract/10.1103/PhysRevLett.129.121104},
   year = {2022}
}

@article{DD2,
  title = {Composition and thermodynamics of nuclear matter with light clusters},
  author = {Typel, S. and R\"opke, G. and Kl\"ahn, T. and Blaschke, D. and Wolter, H. H.},
  journal = {Phys. Rev. C},
  volume = {81},
  issue = {1},
  pages = {015803},
  numpages = {22},
  year = {2010},
  month = {Jan},
  publisher = {American Physical Society},
  doi = {10.1103/PhysRevC.81.015803},
  url = {https://link.aps.org/doi/10.1103/PhysRevC.81.015803}
}

@article{Fortin:2020qin,
    author = "Fortin, M. and Raduta, Adriana R. and Avancini, Sidney and Provid{\^e}ncia, Constan{\c{c}}a",
    title = "{Relativistic hypernuclear compact stars with calibrated equations of state}",
    eprint = "2001.08036",
    archivePrefix = "arXiv",
    primaryClass = "hep-ph",
    doi = "10.1103/PhysRevD.101.034017",
    journal = "Phys. Rev. D",
    volume = "101",
    number = "3",
    pages = "034017",
    year = "2020"
}

@article{Lam2025,
  title = {Axisymmetric stability of neutron stars as extreme rotators in massive scalar-tensor theory},
  author = {Lam, Alan Tsz-Lok and Staykov, Kalin V. and Kuan, Hao-Jui and Doneva, Daniela D. and Yazadjiev, Stoytcho S.},
  journal = {Phys. Rev. D},
  volume = {111},
  issue = {10},
  pages = {104030},
  numpages = {12},
  year = {2025},
  month = {May},
  publisher = {American Physical Society},
  doi = {10.1103/PhysRevD.111.104030},
  url = {https://link.aps.org/doi/10.1103/PhysRevD.111.104030}
}

@article{KEH1,
   author = {Hidemi Komatsu and Yoshiharu Eriguchi and Izumi Hachisu},
   doi = {10.1093/MNRAS/237.2.355},
   issn = {0035-8711},
   issue = {2},
   journal = {Monthly Notices of the Royal Astronomical Society},
   month = {3},
   pages = {355-379},
   publisher = {Oxford Academic},
   title = {Rapidly rotating general relativistic stars – I. Numerical method and its application to uniformly rotating polytropes},
   volume = {237},
   url = {https://dx.doi.org/10.1093/mnras/237.2.355},
   year = {1989}
}

@article{KEH2,
   author = {Hidemi Komatsu and Yoshiharu Eriguchi and Izumi Hachisu and Hidemi Komatsu and Yoshiharu Eriguchi and Izumi Hachisu},
   doi = {10.1093/MNRAS/239.1.153},
   issn = {0035-8711},
   issue = {1},
   journal = {MNRAS},
   month = {7},
   pages = {153-171},
   publisher = {Oxford University Press (OUP)},
   title = {Rapidly rotating general relativistic stars. II - Differentially rotating polytropes},
   volume = {239},
   url = {https://ui.adsabs.harvard.edu/abs/1989MNRAS.239..153K/abstract},
   year = {1989}
}

@article{Stergioulas1995,
   author = {Nikolaos Stergioulas and John L. Friedman},
   doi = {10.1086/175605},
   issn = {0004-637X},
   journal = {The Astrophysical Journal},
   month = {5},
   pages = {306},
   publisher = {American Astronomical Society},
   title = {Comparing models of rapidly rotating relativistic stars constructed by two numerical methods},
   volume = {444},
   url = {https://ui.adsabs.harvard.edu/abs/1995ApJ...444..306S/abstract},
   year = {1995},
}

@Misc{EinsteinToolkit:2025_05,
  requested-for ={EinsteinToolkit},
  author       = {Maxwell Rizzo and others},
  title        = {The {E}instein {T}oolkit},
  month        = may,
  year         = 2025,
  publisher    = {Zenodo},
  version      = {The "Martin D. Kruskal" release, ET\_2025\_05},
  doi          = {10.5281/zenodo.15520463},
  url          = {https://doi.org/10.5281/zenodo.15520463},
}

@article{GRHayL2026,
  title = {Modern, infrastructure-agnostic, extensible library for GRMHD simulations},
  author = {Cupp, Samuel and Werneck, Leonardo R. and Jacques, Terrence Pierre and Tootle, Samuel and Etienne, Zachariah B.},
  journal = {Phys. Rev. D},
  pages = {--},
  year = {2026},
  month = {Jan},
  publisher = {American Physical Society},
  doi = {10.1103/bbtm-31r5},
  url = {https://link.aps.org/doi/10.1103/bbtm-31r5}
}

@article{PPM_method,
title = {The Piecewise Parabolic Method (PPM) for gas-dynamical simulations},
journal = {Journal of Computational Physics},
volume = {54},
number = {1},
pages = {174-201},
year = {1984},
issn = {0021-9991},
doi = {https://doi.org/10.1016/0021-9991(84)90143-8},
author = {Phillip Colella and Paul R Woodward}
}

@book{Book_Riemann_SOLVER,
    author = "Eleuterio F. Toro",
    title = "{Riemann Solvers and Numerical Methods for Fluid Dynamics}",
    publisher = "Springer-Verlag Berlin Heidelberg",
    year = "2009",
    edition ="3rd Edition"
}

@article{Palenzuela:2015dqa,
    author = "Palenzuela, Carlos and Liebling, Steven L. and Neilsen, David and Lehner, Luis and Caballero, O. L. and O'Connor, Evan and Anderson, Matthew",
    title = "{Effects of the microphysical Equation of State in the mergers of magnetized Neutron Stars With Neutrino Cooling}",
    eprint = "1505.01607",
    archivePrefix = "arXiv",
    primaryClass = "gr-qc",
    doi = "10.1103/PhysRevD.92.044045",
    journal = "Phys. Rev. D",
    volume = "92",
    number = "4",
    pages = "044045",
    year = "2015"
}

@article{Alcubierre2003,
   author = {Miguel Alcubierre and Bernd Brügmann and Peter Diener and Michael Koppitz and Denis Pollney and Edward Seidel and Ryoji Takahashi},
   doi = {10.1103/PhysRevD.67.084023},
   issn = {15502368},
   issue = {8},
   journal = {Physical Review D},
   month = {4},
   pages = {084023},
   publisher = {American Physical Society},
   title = {Gauge conditions for long-term numerical black hole evolutions without excision},
   volume = {67},
   url = {https://journals.aps.org/prd/abstract/10.1103/PhysRevD.67.084023},
   year = {2003}
}

@article{Bona1995,
  title = {New Formalism for Numerical Relativity},
  author = {Bona, Carles and Mass\'o, Joan and Seidel, Edward and Stela, Joan},
  journal = {Phys. Rev. Lett.},
  volume = {75},
  issue = {4},
  pages = {600--603},
  numpages = {0},
  year = {1995},
  month = {Jul},
  publisher = {American Physical Society},
  doi = {10.1103/PhysRevLett.75.600},
  url = {https://link.aps.org/doi/10.1103/PhysRevLett.75.600}
}

@inbook{KO_dissipation,
publisher = {John Wiley \& Sons, Ltd},
isbn = {9781118548448},
author = { Gustafsson, B. and  Kreiss, H. and  Oliger, J},
title = {References},
booktitle = {Time‐Dependent Problems and Difference Methods},
chapter = {},
pages = {499-505},
doi = {https://doi.org/10.1002/9781118548448.refs},
year = {2013}
}

@article{ROZWADOWSKA2021101498,
	author = {Karolina Rozwadowska and Francesco Vissani and Enrico Cappellaro},
	doi = {https://doi.org/10.1016/j.newast.2020.101498},
	issn = {1384-1076},
	journal = {New Astronomy},
	pages = {101498},
	title = {On the rate of core collapse supernovae in the milky way},
	url = {https://www.sciencedirect.com/science/article/pii/S138410762030289X},
	volume = {83},
	year = {2021}}

@inbook{Cerda-Duran2018,
	address = {Cham},
	author = {Cerda-Duran, Pablo and Elias-Rosa, Nancy},
	booktitle = {The Physics and Astrophysics of Neutron Stars},
	doi = {10.1007/978-3-319-97616-7_1},
	editor = {Rezzolla, Luciano and Pizzochero, Pierre and Jones, David Ian and Rea, Nanda and Vida{\~{n}}a, Isaac},
	isbn = {978-3-319-97616-7},
	pages = {1--56},
	publisher = {Springer International Publishing},
	title = {Neutron Stars Formation and Core Collapse Supernovae},
	url = {https://doi.org/10.1007/978-3-319-97616-7_1},
	year = {2018}}

@article{Burrows2023,
   author = {Adam Burrows and David Vartanyan and Tianshu Wang},
   doi = {10.3847/1538-4357/ACFC1C},
   issn = {0004-637X},
   issue = {2},
   journal = {The Astrophysical Journal},
   month = {10},
   pages = {68},
   publisher = {IOP Publishing},
   title = {Black Hole Formation Accompanied by the Supernova Explosion of a 40 M ⊙ Progenitor Star},
   volume = {957},
   url = {https://iopscience.iop.org/article/10.3847/1538-4357/acfc1c https://iopscience.iop.org/article/10.3847/1538-4357/acfc1c/meta},
   year = {2023}
}

@article{Zhang2008,
   author = {Weiqun Zhang and S. E. Woosley and A. Heger},
   doi = {10.1086/526404},
   issn = {0004-637X},
   issue = {1},
   journal = {The Astrophysical Journal},
   month = {5},
   pages = {639},
   publisher = {IOP Publishing},
   title = {Fallback and Black Hole Production in Massive Stars},
   volume = {679},
   url = {https://iopscience.iop.org/article/10.1086/526404 https://iopscience.iop.org/article/10.1086/526404/meta},
   year = {2008}
}

@article{Igoshev2022,
   author = {Andrei P. Igoshev and Anastasia Frantsuzova and Konstantinos N. Gourgouliatos and Savina Tsichli and Lydia Konstantinou and Sergei B. Popov},
   doi = {10.1093/MNRAS/STAC1648},
   issn = {0035-8711},
   issue = {3},
   journal = {Monthly Notices of the Royal Astronomical Society},
   month = {6},
   pages = {4606-4619},
   publisher = {Oxford Academic},
   title = {Initial periods and magnetic fields of neutron stars},
   volume = {514},
   url = {https://dx.doi.org/10.1093/mnras/stac1648},
   year = {2022}
}

@article{Ozel2016,
   author = {Feryal Özel and Paulo Freire},
   doi = {10.1146/ANNUREV-ASTRO-081915-023322/CITE/REFWORKS},
   issn = {00664146},
   issue = {Volume 54, 2016},
   journal = {Annual Review of Astronomy and Astrophysics},
   month = {9},
   pages = {401-440},
   publisher = {Annual Reviews Inc.},
   title = {Masses, radii, and the equation of state of neutron stars},
   volume = {54},
   url = {https://www.annualreviews.org/content/journals/10.1146/annurev-astro-081915-023322},
   year = {2016}
}

@article{Thornton2013,
	author = {D. Thornton and B. Stappers and M. Bailes and B. Barsdell and S. Bates and N. D. R. Bhat and M. Burgay and S. Burke-Spolaor and D. J. Champion and P. Coster and N. D'Amico and A. Jameson and S. Johnston and M. Keith and M. Kramer and L. Levin and S. Milia and C. Ng and A. Possenti and W. van Straten},
	doi = {10.1126/science.1236789},
	eprint = {https://www.science.org/doi/pdf/10.1126/science.1236789},
	journal = {Science},
	number = {6141},
	pages = {53-56},
	title = {A Population of Fast Radio Bursts at Cosmological Distances},
	url = {https://www.science.org/doi/abs/10.1126/science.1236789},
	volume = {341},
	year = {2013}}

@article{Yamasaki2023,
   author = {Shotaro Yamasaki and Tomotsugu Goto and Chih Teng Ling and Tetsuya Hashimoto},
   doi = {10.1093/MNRAS/STAD3844},
   issn = {0035-8711},
   issue = {4},
   journal = {Monthly Notices of the Royal Astronomical Society},
   month = {12},
   pages = {11158-11166},
   publisher = {Oxford Academic},
   title = {The true fraction of repeating fast radio bursts revealed through CHIME source count evolution},
   volume = {527},
   url = {https://dx.doi.org/10.1093/mnras/stad3844},
   year = {2023}
}

@article{Lorimer2008,
   author = {Duncan R. Lorimer},
   doi = {10.12942/LRR-2008-8/TABLES/4},
   issn = {14338351},
   issue = {1},
   journal = {Living Reviews in Relativity},
   month = {11},
   pages = {8-},
   publisher = {Albert Einstein Institut},
   title = {Binary and millisecond pulsars},
   volume = {11},
   url = {https://link.springer.com/article/10.12942/lrr-2008-8},
   year = {2008}
}

@article{Swihart2022,
   author = {Samuel J. Swihart and Jay Strader and Laura Chomiuk and Elias Aydi and Kirill V. Sokolovsky and Paul S. Ray and Matthew Kerr},
   doi = {10.3847/1538-4357/ACA2AC},
   issn = {0004-637X},
   issue = {2},
   journal = {The Astrophysical Journal},
   month = {12},
   pages = {199},
   publisher = {IOP Publishing},
   title = {A New Flaring Black Widow Candidate and Demographics of Black Widow Millisecond Pulsars in the Galactic Field},
   volume = {941},
   url = {https://iopscience.iop.org/article/10.3847/1538-4357/aca2ac https://iopscience.iop.org/article/10.3847/1538-4357/aca2ac/meta},
   year = {2022}
}

@article{Koljonen2025,
   author = {Karri I. I. Koljonen and Manuel Linares},
   doi = {10.3847/1538-4357/AE08A5},
   issn = {0004-637X},
   issue = {1},
   journal = {The Astrophysical Journal},
   month = {11},
   pages = {8},
   publisher = {IOP Publishing},
   title = {SpiderCat: A Catalog of Compact Binary Millisecond Pulsars},
   volume = {994},
   url = {https://iopscience.iop.org/article/10.3847/1538-4357/ae08a5 https://iopscience.iop.org/article/10.3847/1538-4357/ae08a5/meta},
   year = {2025}
}

@article{Adams2013,
   author = {Scott M. Adams and C. S. Kochanek and John F. Beacom and Mark R. Vagins and K. Z. Stanek},
   doi = {10.1088/0004-637X/778/2/164},
   issn = {0004-637X},
   issue = {2},
   journal = {The Astrophysical Journal},
   month = {11},
   pages = {164},
   publisher = {IOP Publishing},
   title = {OBSERVING THE NEXT GALACTIC SUPERNOVA},
   volume = {778},
   url = {https://iopscience.iop.org/article/10.1088/0004-637X/778/2/164 https://iopscience.iop.org/article/10.1088/0004-637X/778/2/164/meta},
   year = {2013}
}

\section*{Supplemental material}

\section{Modeling the supramassive NS progenitor in MSTT.}

With the scalar-field mass $m_\Phi$ specified, the MSTT parameters $(\alpha_0,\beta_0)$ can be constrained by requiring the theory to reproduce observed NS properties. Since the theory must simultaneously support the observed masses and rotation rates of known pulsars, we consider a set of parameters capable of reproducing the observed properties of the black widow pulsar PSR J0952-0607 \cite{Romani2022}. This is among the most massive and fastest spinning NS known, with a gravitational mass of $2.35\pm0.11\,M_\odot$ and a spin frequency of $707$ Hz \cite{Romani2025}. By reproducing the extreme properties of PSR J0952-0607 for a given EoS and theory parameters, our model will inherently accommodate the entire, less extreme pulsar population.

The spin frequency is measured independently of the gravitational theory for PSR J0952-0607. The mass measurement of the NS is derived from the motion and heating of its subsolar mass companion star, obtained through purely Newtonian methods in which the gravity theory enters only through the gravitational constant $G$ \cite{Romani2022}. As such, deviations in the gravitational mass associated with MSTT are negligible compared to the systematic errors of the measurement and can be treated as independent of the theory of gravity. 
The mass of the NS and its spin frequency together fix  quantities such as the star's angular momentum, which intrinsically depends on both the theory of gravity and the EoS. Not all EoS/theory parameters are able to reproduce the properties of PSR J0952-0607. In particular, in GR, most hyperonic EoS are unable to reach such high masses. This problem is not present in MSTT, since the maximum mass of a NS can increase due to scalarization for appropriate values of ($m_\Phi,\alpha_0,\beta_0$).

\begin{figure}[t]
\includegraphics[width=0.95\linewidth]{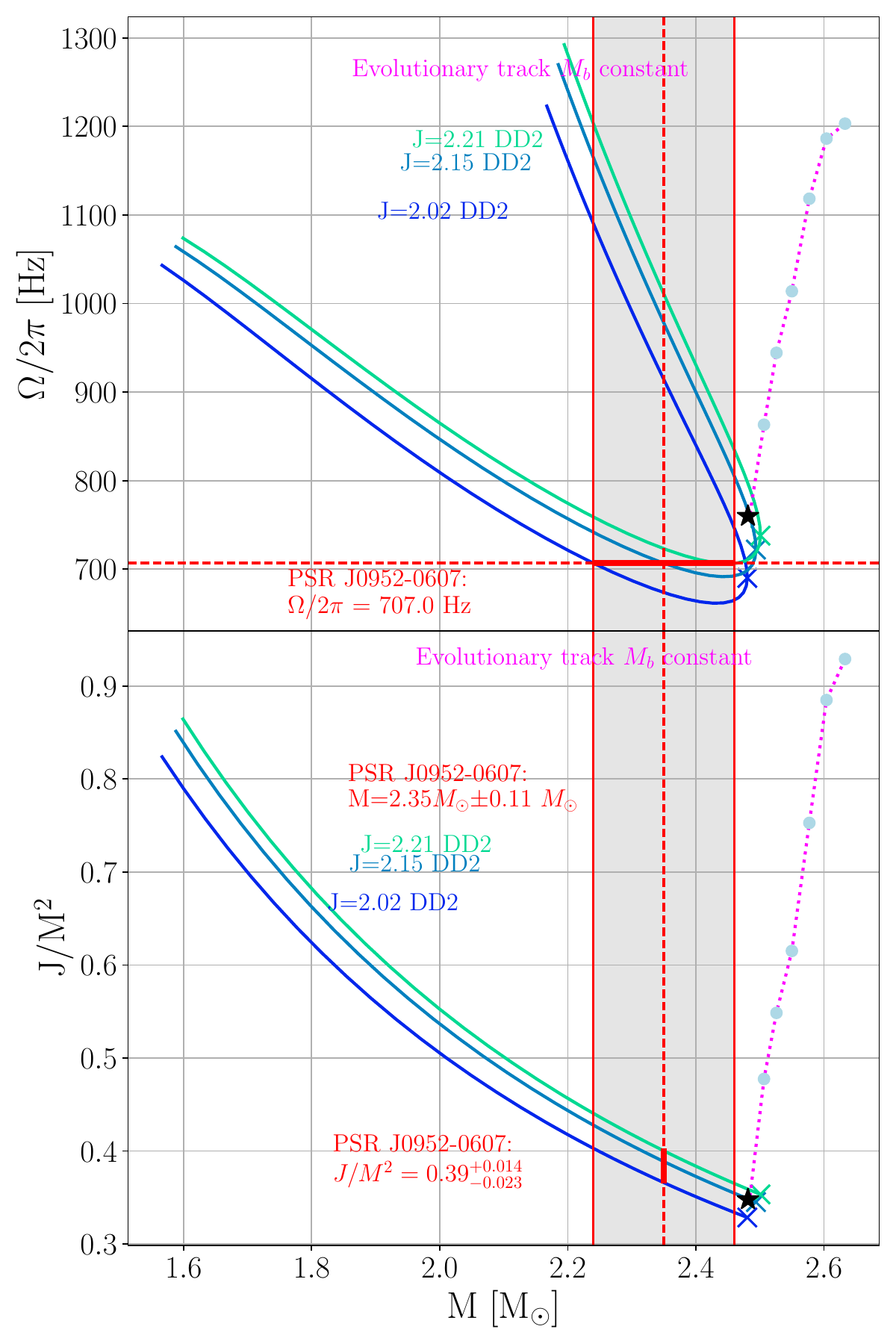}
\caption{Parameter space for pulsar PSR J0952-0607 in MSTT. Top and bottom panels show, respectively, the star's spin frequency and spin parameter $\chi=J/M^2$ as a function of the gravitational mass. The solid colored lines represent constant angular momentum sequences. The crosses mark the maximum gravitational mass allowed for each sequence. The shaded region and red lines give the measured mass and spin frequency of PSR J0952-0607 and their uncertainties \cite{Romani2025}. The black star represents the supramassive progenitor we evolve while the dotted magenta lines represent the evolutionary track it follows as it spins-down.}
\label{fig:mass_freq_PSR}
\end{figure}

Fig.~\ref{fig:mass_freq_PSR} displays the spin frequency $\Omega/2\pi$ (top panel) and the dimensionless spin parameter $\chi=J/M^2$ (bottom panel), both as a function of the gravitational mass $M$. The dashed vertical line represents the central value of the mass reported in \cite{Romani2025}, while the shaded region represents the uncertainty in the mass measurement. The dashed horizontal line in the top panel marks the observed spin frequency. The solid curves represent sequences of constant angular momentum $J$ computed for the fixed set of MSTT parameters: $\alpha_0=10^{-2}$, $\beta_0=-10$ and $m_\Phi=10^{-11}$ eV$/c^2$ with the cold $\beta$-equilibrium DD2 EoS \cite{DD2}, in its hyperonic version \cite{Fortin:2020qin}.  For each constant-$J$ sequence, we identify the maximum mass, marked by a cross in Fig.~\ref{fig:mass_freq_PSR}. According to the turning-point criterion \cite{Friedman1988}, which has been validated for MSTT \cite{Lam2025,OlveraM:2026whf}, configurations beyond this maximum mass point along the sequence are unstable to axisymmetric perturbations. The dotted magenta line represents the evolutionary track a supramassive star with baryon mass $M_b$ follows as it spins-down and migrates from one constant-$J$ sequence to another with a lower value of $J$ until it becomes unstable, marked with a black star in the figure.

The intersection of the uncertainty region with the constant angular momentum sequences yields a spin parameter $\chi^{\rm PSR}=0.39_{-0.023}^{+0.014}$ for pulsar PSR J0952-0607 in our MSTT-DD2 model with an angular momentum of $J^{\rm PSR}=2.15^{+0.06}_{-0.13} M_\odot^{2}$. In this case, PSR J0952-0607 serves only to fix an angular momentum which can be achieved in a realistic astrophysical scenario. The progenitor we evolve is a supramassive NS configuration (marked with a black star in Fig. \ref{fig:mass_freq_PSR}) which lies along the unstable branch of the constant-$J$ sequence that corresponds to the central value of $\chi^{\rm PSR}$ and $J^{\rm PSR}$ for PSR J0952-0607.  Our progenitor has a gravitational mass $M_\star$ larger than that for PSR J0952-0607 and it also spins faster. 

Astrophysically, it represents a NS born with baryon mass $M_b$, angular momentum $J_\star^0$, and spin frequency $\Omega_*^0/2\pi$. As it spins down, it loses angular momentum, until it crosses the turning point at $J^{\rm PSR}$ and gravitational mass $M_\star$ and collapses to a BH. We consider that $M_\star$ is $0.5\%$ less massive than the maximum mass allowed by the sequence. We ensure $M_{\rm{\star}}>M_{\rm{TOV}}$, where $M_{\rm{TOV}}$ is the non-rotating maximum mass of the corresponding configuration. The same analysis is  repeated for the light scalar field mass model, with  $\alpha_0=10^{-2}$, $\beta_0=-6$ and $m_\Phi=3.3\times10^{-12}$ eV$/c^2$ to determine the initial collapsing model. The values of the relevant parameters for the two models are reported in Table~\ref{tab:models}.

\begin{table}[t]
    \centering
    \caption{Supramassive NS models employed in this work. Both models employ the DD2 EoS in its hyperonic version~\cite{Fortin:2020qin}. Shown in the table are the MSTT parameters $\alpha_0$, $\beta_0$ and $m_\Phi$, along with the gravitational and baryon mass of the star $M_\star$ and $M_b$, and its dimensionless spin parameter $\chi$.     }
    \begin{tabular}{c|c|c|c|c|c|c}
    \hline
    Model &  $\alpha_0$ & $\beta_0$ & $m_\Phi$ [eV] & $M_\star$ [$M_\odot$] & $M_b$ [$M_\odot$] & $\chi$ \\
    \hline
        Light &  $10^{-2}$ & -6 & $3.3\times10^{-12}$ & 2.296 & 2.697 & 0.349 \\
        Heavy &  $10^{-2}$ & -10 & $10^{-11}$          & 2.527 & 3.012    & 0.346\\
        \hline
    \end{tabular}
    \label{tab:models}
\end{table}

\section{Numerical details}\label{sec:numerics}

Once the angular momentum of the NS progenitor has been identified, we construct rapidly rotating configurations using the Komatsu–Eriguchi–Hachisu self-consistent field method~\cite{KEH1,KEH2} with the \texttt{RNS} code \cite{Stergioulas1995} modified for MSTT~\cite{Doneva2013}. These models are then used as initial data for the simulations.  We apply a radial velocity perturbation $\delta v= 3\times10^{-4} r/R_e$ to trigger collapse (here $R_e$ is the equatorial circumferential radius).

The simulations are performed using the \textsc{Einstein Toolkit} \cite{EinsteinToolkit:2025_05}, with the same setup as  in \cite{OlveraM:2026whf}, to which the interested reader is addressed for details. The scalar field extraction is performed at $5\lambda_\Phi$, where $\lambda_\Phi$ is the Compton wavelength associated with the mass of the scalar field. This corresponds to $\sim 20$ km for the heavy case and $\sim 60$ km for the light case.

The spacetime evolution is performed using the \texttt{JBSSN} module described in our previous work \cite{OlveraM:2026whf}. \texttt{JBSSN} consists of a Baumgarte-Shapiro-Shibata-Nakamura (BSSN) formulation of the Einstein-MSTT system of equations in the Jordan frame: 
\begin{equation}\label{eq:EinsteinEqs}
    \begin{aligned}
       G_{\mu\nu} = \frac{8\pi G}{\Phi}\left( T^{\rm matter}_{\mu\nu} + T^{\rm SF}_{\mu\nu}\right)\,,
    \end{aligned}
\end{equation}
\begin{equation}\label{eq:SF}
    \begin{aligned}
        \Box \Phi = \frac{1}{2\omega(\Phi)+3}\left[ 8\pi T - \frac{d\omega}{d\Phi}\nabla_\alpha\Phi\nabla^\alpha\Phi \right. \\ +  \Phi^3\frac{d}{d\Phi} \left.\left(\frac{U(\Phi)}{\Phi^2} \right)\right]\,,
    \end{aligned}
\end{equation}
where $\Box$ is the d'Alembertian operator and $T$ is the trace of the energy-momentum tensor, with the effective energy-momentum tensor of the scalar field, $ T^{\rm SF}_{\mu\nu}$ given by
\begin{equation}\label{eq:SF_T}
    \begin{aligned}
      8 \pi T^{\rm SF}_{\mu\nu} = \frac{\omega(\Phi)}{\Phi} &\left( \nabla_\mu \Phi \nabla_\nu \Phi  -\frac{1}{2} g_{\mu\nu} \nabla_\alpha\Phi\nabla^\alpha\Phi  \right) \\+ & \nabla_\mu\nabla_\nu \Phi -g_{\mu\nu} \Box \Phi-\frac{1}{2} g_{\mu\nu} U(\Phi)\, .
    \end{aligned}
\end{equation}

\texttt{JBSSN} evolves the scalar field equations and the spacetime BSSN-like system using the 1+log slicing condition \cite{Bona1995} with the Gamma-driver condition \cite{Alcubierre2003}. By performing the spacetime evolution in the Jordan frame, the hydrodynamic system is reduced to the same evolution equations employed in GR. Hydrodynamic evolution is performed using the \texttt{GRHayL} module \cite{GRHayL2026}, with a piecewise-parabolic-method for cell-reconstruction \cite{PPM_method} and the HLLE solver for shock-capturing of Riemmann problems \cite{Book_Riemann_SOLVER}. The conservative-to-primitive variable conversion follows the method of~\cite{Palenzuela:2015dqa}. All simulations employ an atmosphere floor of $10^{-10}$ times the initial central density of the NS and Kreiss-Oliger dissipation with $\epsilon^{\rm KO}=0.025$ \cite{KO_dissipation}.    

The numerical grid extends up to $\sim\pm945\,\rm km$ in every Cartesian direction and is composed by five initial refinement levels with a finest resolution of $\sim 296\,\rm m$. As the simulation progresses, three additional refinement levels are added in order to track the appearance of the BH's apparent horizon. 

\section{Astrophysical rates}\label{sec:rates}

This section discusses our estimates for the Galactic rates of NS collapse to BH through three main astrophysical formation channels: BH formation through failed CCSN, the collapse of supramassive NSs, and accretion-induced collapse. Although our simulations cover only the second channel, we also estimate the rates for the other channels for completeness.

\begin{table*}[t]
\centering
\caption{Estimated Galactic rates for the NS collapse channels considered in this work.}
\label{tab:rates}
\begin{tabular}{c|c|c|c}
\hline
Channel & Parent population & Rate $(100\,{\rm yr})^{-1}$ & Constraint \\
\hline
Failed CCSN & Massive stars & $0.05$--$1.3$ & Observed failed-CCSN fraction \\
Supramassive & CCSNe & $<6\times10^{-2}$ & FRB (blitzar) limit \\
Black widow & Recycled MSPs & $f_{\rm coll}^{\rm BW}\times10^{-5}$ & Spider population \\
NS-AIC & Recycled MSPs & $f_{\rm coll}^{\rm BW}\times10^{-5} \lesssim R_{\rm NS-AIC} < 10^{-4}$ &  MSP recycling rate \\
\hline
\end{tabular}
\end{table*}

\subsection{BH formation by failed CCSN}

A conservative estimate of the Galactic CCSN rate inferred from observed explosions yields $R_{\rm CCSN}=1.63\pm0.46\,(100\,\mathrm{yr})^{-1}$, that is, one CCSN event every $61^{+24}_{-14}$ years \cite{ROZWADOWSKA2021101498}. For a proto-neutron star (PNS), thermal pressure support on the core results in the maximum allowed mass (or the threshold for instability) rising by up to $\sim$$25\%$ compared with a stationary cold sequence in equilibrium \cite{OConnor2011}. A BH can form either via a failed supernova, where the progenitor cannot launch a strong enough shock, or through significant fallback accretion following an explosion. It is expected that in both cases the progenitor is a hot, lepton-rich PNS, which radiates neutrinos until collapsing inside its event horizon \cite{Sukhbold_2016}, rather than the cold collapsing configurations studied here. PNS formation in MSTT has been shown to strongly excite the scalar sector in some cases, producing scalar radiation \cite{Rosca-Mead2020,Sperhake:2017itk}. 

The rate of failed explosions is tightly tied to the characteristics of the progenitor massive star \cite{Sukhbold_2016,OConnor2011,Cerda-Duran2018} and recent searches put a bound on the fraction of failed CCSNe of around $f=16^{+23}_{-12}\%$ with $90\%$ confidence \cite{Neustadt2021}. Therefore, the rate of BH formation from this first channel is approximately $f/(1-f)\,R_{\rm CCSN}\gtrsim 0.31\,(100\,\mathrm{yr})^{-1}$, where the factor $1/(1-f)$ accounts for the failed fraction being defined over all core collapses while $R_{\rm CCSN}$ counts successful explosions. Combining the extremes of both uncertainties gives a range $0.05$--$1.3\,(100\,\mathrm{yr})^{-1}$. Moreover, since fallback-driven BH formation involves an explosion \cite{Burrows2023,Zhang2008}, it escapes the optical disappearance searches from \cite{Neustadt2021} and as a result, our estimate is a lower bound.

PNS that survive this phase cool and contract through a quasi-stationary Kelvin-Helmholtz phase, lasting around tens of seconds \cite{Pons1999,Burrows:1986me} and settle into cold, stable neutron stars. Equivalently,  the birth rate of long-lived neutron stars produced by successful CCSNe is $R_{\rm CCSN}$.

The subsequent PNS evolution branches into the remaining channels: neutron stars born massive and rapidly rotating, which collapse once magnetic-dipole braking removes their centrifugal support, and neutron stars in X-ray binaries that are spun up by accretion, the path leading to accretion-induced collapse and to black-widow systems such as PSR~J0952$-$0607.

\subsection{Supramassive NS collapse}

The spin-down channel requires a NS born with a mass above the non-rotating maximum mass but still rotating fast enough so that centrifugal support renders it metastable \cite{Falcke2014,Ravi2014}. Such a supramassive pulsar collapses once magnetic braking has removed sufficient angular momentum, after a spin-down time of $10^{2}$--$10^{6}$ yr, with this timescale being determined by how fast the period of the star is at birth and how massive it is \cite{Falcke2014}.  Assuming that supramassive NSs are produced mainly during CCSNe, the supramassive birth rate is given by $R_{\rm sup}=f_{\rm sup}\,R_{\rm CCSN}$. The lower bound of the birth fraction $f_{\rm sup}$ is highly dependent on the EoS since it requires both a birth mass above $M_{\rm TOV}$ and a birth period of a few milliseconds. The inferred pulsar birth periods cluster at tens to hundreds of milliseconds \cite{Igoshev2022} and birth masses near  $1.3$--$1.4\,M_\odot$ \cite{Ozel2016}.  

A ceiling on $f_{\rm sup}$ follows from the fast radio burst (FRB) population. The collapse of a magnetized supramassive pulsar is expected to be accompanied by a blitzar, which in principle could lead to a FRB \cite{Falcke2014}.  Assuming that each spin-down induced collapse yields an observable FRB, the whole FRB population \cite{Thornton2013} can be explained with only $\sim3\%$ of the CCSNe rate within $z\lesssim1$ resulting in supramassive remnants. If supramassive NS births were more frequent than this, the FRB population would be overproduced \cite{Falcke2014}.   Assuming that the same rate for $z=1$ applies to the present-day Milky Way galaxy, this sets the maximum allowed value $f_{\rm sup}\lesssim3\times10^{-2}$, i.e. $R_{\rm sup}\lesssim6\times10^{-2}\,(100\,\mathrm{yr})^{-1}$. This upper limit would be relaxed if a fraction of such collapses do not present radio emission. It is important to mention that recent works also show that $\sim50\%$ of the FRB population are repeating sources \cite{Yamasaki2023}, reducing our estimate by a factor of about two, since collapsing sources only display a single burst.

\subsection{Accretion-induced collapse }

The accretion-driven channels follow from the initial population $R_{\rm CCSN}$ scaled by the Galactic NS recycling  rate. The formation rate of galactic milisecond pulsars (MSPs) is given by observational estimates at $R_{\rm MSP}\sim 10^{-4}\,(100\,\mathrm{yr})^{-1}$ \cite{Lorimer2008}.

 Accretion-induced collapse of a recycled NS (NS-AIC) in a low-mass X-ray binary \cite{Dermer2006} is estimated as $R_{\rm MSP}$ times the fraction of recycled stars born close enough to the maximum mass such that the accreted mass drives them over the threshold. The observed recycled-pulsar mass distribution is typically centered around $1.54M_\odot$ \cite{Antoniadis:2016hxz,Ozel2016}. While EoS uncertainties and systematics in population studies make an exact count difficult, black widow systems can be used to constrain the collapse rate for this specific channel. 
 
For the black-widow endpoint, spider systems constitute $15$--$20\%$ of the recycled MSPs~\cite{Swihart2022}. According to the most recent catalog of spider MSPs \cite{Koljonen2025}, out of the confirmed 82 entries, 50 ($\sim60\%$) of them are categorized as black widow systems ($9-12\%$ of the total galactic MSP population). As a result, the black widow collapse rate is estimated as $R_{\rm BW}\sim f^{\rm BW}_{\rm coll}\times 10^{-5}\,(100\,\mathrm{yr})^{-1}$, where $f^{\rm BW}_{\rm coll}$ represents the fraction of Galactic black widows that exceed the maximum stable mass and collapse to a BH, that is, $f^{\rm BW}_{\rm coll}=N_{\rm BW}^{\rm Collapse}/N_{\rm BW}$.  Since current observations do not constrain the fraction of black widows that ultimately exceed the maximum stable mass, $0 \leq f^{\rm BW}_{\rm coll} \leq 1$.  

With this we can bound the collapse rate of accretion-induced channels. Since BW collapse is already a particular subset of endpoints for NS-AIC, $R_{\rm BW} \lesssim R_{\rm NS-AIC}$, and since not al MSPs collapse to a BH, then $R_{\rm NS-AIC}<R_{\rm MSP}$. As a result, $f^{\rm BW}_{\rm coll}\times 10^{-5}\,(100\,\mathrm{yr})^{-1}\lesssim R_{\rm NS-AIC} <  10^{-4}\,(100\,\mathrm{yr})^{-1}$, which already shows  this channel is subdominant to the other two.

\subsection{Comparison between BH formation channels}
For sources within 8.5 kpc from Earth, we rescale each Galactic rate by the fraction of the corresponding parent population enclosed within this volume. For the star formation channels that follow from the Galactic CCSN rate (failed CCSNe, supramassive pulsars) this fraction is $f_{\rm geo}(8.5\rm\,kpc)\sim0.45$, as can be seen from the cumulative CCSN distance distribution of \cite{Adams2013}. 

 The rate within 8.5 kpc is dominated by the failed-CCSN channel at $\gtrsim 0.16_{-0.13}^{+0.49}\,(100\,\mathrm{yr})^{-1}$, corresponding to one descalarization burst every $\sim600$ yr (range $150$--$4000$ yr).  The spin-down of supramassive stars remains upper bounded by $R_{\rm{sup}}\lesssim3\times10^{-2}(100\rm{yr})^{-1}$, while NS-AIC is bounded by $R_{\rm BW}\lesssim R_{\rm NS-AIC}<10^{-4}(100\rm{yr})^{-1}$. A summary of each channel, as well as of its main uncertainties, is shown in Table \ref{tab:rates}.

Although failed CCSNe dominate the event rate, the scalar signal originates from a hot PNS embedded within a dense stellar envelope, which is not necessarily scalarized. By contrast, supramassive and accretion-driven collapse occur from cold equilibrium neutron stars similar to those studied here and as a result of our modelling, these formation channels are always accompanied by a descalarization burst. Quantifying this difference requires dedicated simulations.

\end{document}